\documentclass[%
 reprint,
superscriptaddress,
nofootinbib,
 amsmath,amssymb,
 aps
]{revtex4-2}

\usepackage[T1]{fontenc}
\usepackage{blindtext}
\usepackage[utf8]{inputenc}
\usepackage{graphicx}
\usepackage{subcaption}
\usepackage{dcolumn}
\usepackage{bm}
\usepackage{epsfig}
\usepackage{amsfonts}
\usepackage{amsmath}
\usepackage{amssymb}
\usepackage[usenames]{color}
\usepackage[dvipsnames]{xcolor}
\usepackage{float}
\usepackage{caption}
\usepackage{tabularx,ragged2e,booktabs}
\newcolumntype{C}{>{\Centering\arraybackslash}X}
\usepackage{multirow} 
\usepackage{tensor}
\usepackage{titlesec}
\usepackage[shortlabels]{enumitem}
\usepackage[normalem]{ulem}
\usepackage{fixltx2e}
\usepackage{url}

\usepackage[colorlinks=true, citecolor=blue, linkcolor=blue, urlcolor=blue]{hyperref}
\begin{document}

\title{\boldmath \textbf{Primordial magnetic non-Gaussianity with generic vacua and detection prospects in CMB spectral distortions}}

\author{Arko Bhaumik}
 \email{arkobhaumik12@gmail.com}
 \affiliation{Physics and Applied Mathematics Unit, Indian Statistical Institute,\\203, B.T. Road, Kolkata 700 108, India}
\author{Supratik Pal}
 \email{supratik@isical.ac.in}
 \affiliation{Physics and Applied Mathematics Unit, Indian Statistical Institute,\\203, B.T. Road, Kolkata 700 108, India}
 \affiliation{Technology Innovation Hub on Data Science, Big Data Analytics and Data Curation,\\
Indian Statistical Institute, 203, B.T. Road, Kolkata 700 108, India}

\begin{abstract}
Assuming a slow-roll inflationary model where conformal invariance of the Maxwell action is broken via a non-minimal kinetic coupling term, we investigate the non-Gaussian three-point cross-correlation function between the primordial curvature perturbation and the primordial magnetic field, under a fairly general choice of initial vacua for both the scalar and the gauge field sectors. Among the possible triangular configurations of the resulting cross-bispectrum, we find that the squeezed limit leads to local-type non-Gaussianity allowing a product form decomposition in terms of the scalar and magnetic power spectra, which is a generic result independent of any specific choice of the initial states. We subsequently explore its detection prospects in the CMB via correlations between pre-recombination $\mu$-type spectral distortions and temperature anisotropies, sourced by such a primordial cross-correlation. Our analysis with several proposed next-generation CMB missions forecasts a low value of the signal-to-noise ratio (SNR) for the $\mu T$ spectrum if both the vacua are assumed to be pure Bunch-Davies. On the contrary, the SNR may be enhanced significantly for non-Bunch-Davies initial states for the magnetic sector within allowed bounds from current CMB data.
\end{abstract}

\maketitle

\flushbottom

\section{Introduction} \label{sec:intro}

Over the past few decades, the presence of very weak magnetic fields coherent across megaparsec (Mpc) scales in the observable Universe has been indicated by a growing body of observational evidence. Although we are yet to directly observe the effects of such a magnetic field on the cosmic microwave background (CMB), available datasets \cite{faraday_pmf_bound_1,faraday_pmf_bound_2,faraday_pmf_bound_3,faraday_pmf_bound_4,cmb_pmf_bound_1,cmb_pmf_bound_2,cmb_pmf_bound_3} place an upper bound of the order $B_0\lesssim10^{-9}$ G on the strength of any such large-scale magnetic field. 
On the other hand, observations of secondary GeV-scale gamma-ray cascades associated with distant TeV-blazars imply a lower limit of around $B_0\gtrsim10^{-16}$ G \cite{blazar_bound_1,blazar_bound_2,blazar_bound_3}. Unlike the micro-Gauss strength fields pervading galaxies and galaxy clusters that may be explained by dynamo amplification of pre-existing seed fields \cite{dynamo_rev_1,dynamo_rev_2}, the magnetic fields in the intergalactic void are harder to explain on the basis of astrophysical processes alone \cite{igmf_astro}, thereby hinting at a potentially primordial origin. Inflation is capable of producing such large-scale coherent magnetic fields by a simple prescription where the conformal symmetry of the electromagnetic (EM) action is violated via direct coupling to an auxiliary scalar field, which may also act as the inflaton itself \cite{ratra,martin_yokoyama}. Such a model can generate a nearly scale-invariant spectrum of stochastic primordial magnetic fields (PMF) having the required amplitude on Mpc scales at the present day. Despite limitations such as the strong coupling and back reaction problems as well as possible alternatives (the interested reader may refer to \cite{Kandus:2010nw,problems_2}  for a detailed review), this scenario is widely accepted as the most promising one so far. 

On the other hand, such a scenario may also lead to higher order non-Gaussian correlations between the PMFs and the inflationary metric perturbations \cite{Caldwell_2011,Caldwell_Motta,Jain_Sloth,Jain_grav}, which may leave various observable signatures through the subsequent cosmic epochs. To several next-generation CMB missions, the cross power spectrum $C_\ell^{\mu T}$ between $\mu$-type spectral distortions and temperature anisotropies of the CMB is of particular interest. Such distortions in the CMB blackbody spectrum may typically be caused by an injection of energy into the pre-recombination photon-baryon fluid, which is expected to distort a Planckian distribution into a Bose-Einstein distribution with non-zero chemical potential. This phenomenon  occurs much before recombination, between the initial and final redshifts of $z_\mu^i\sim2.0\times10^6$ and $z_\mu^f\sim5.0\times10^4$ respectively. Prior to this window, at $z\gg z_\mu^i$, photon non-conserving processes like double Compton scattering ($e^-+\gamma\to e^-+2\gamma$) can efficiently redistribute any excess energy injected into the ideal photon gas of temperature $T_1$, and restore thermal equilibrium by re-attaining a Planckian spectrum with a new temperature $T_2$. But for $z\lesssim z_\mu^i$, such processes lose their efficiency, and the new equilibrium needs to be attained with a conserved photon number via the dominant elastic Compton scattering process. This leads to a final Bose-Einstein distribution function given by $n(x)=\left[\textrm{exp}(x+\mu)-1\right]^{-1}$ with $x=h\nu/k_BT$, that is characterized by both a final temperature $T_2$ and a chemical potential $\mu$. The efficiency of this photon-conserving pathway holds until $z\sim z_\mu^f$, after which energy injections during more recent epochs give rise to $y$-distortion effects, which are distinct from the $\mu$-distortion signatures. Possible cosmological sources of spectral distortions have been studied extensively in the literature, which include but are not limited to unstable massive relics \cite{PhysRevLett.70.2661}, pre-recombination dissipation of small-scale PMFs \cite{Jedamzik:1999bm}, primordial scalar non-Gaussianity \cite{newwind,Ganc:2012,Chluba:2016aln,CMB-S4:2023zem}, multiple inflaton scenarios \cite{Bae:2017tll}, enhanced small scale power and primordial black holes \cite{Chluba:2012we,Ozsoy:2021qrg}, axion monodromy inflation \cite{Henriquez-Ortiz:2022ulz}, etc. For a detailed exposition of the major sources of spectral distortions within the $\Lambda$CDM paradigm in particular, we refer the interested reader to \cite{Chluba:2016bvg} and the references therein. Among several such possible origins, the non-Gaussian three-point cross-correlation between the primordial curvature perturbation and PMF turns out to be an interesting candidate that may source a possible $C_\ell^{\mu T}$ correlation, whose resulting signal might be of particular interest to future CMB missions \cite{ganc_sloth}.

So far, primordial cross-correlations between metric perturbations and gauge fields have mostly been studied in the literature under the assumption of standard Bunch-Davies (BD) conditions, which assert an initial Minkowskian vacuum for the inflationary modes when they are deep within the horizon \cite{Chernikov:1968zm,Bunch:1978yq}. However, apart from simplicity in perturbation calculations, there is no compelling reason to choose this specific vacuum a priori. Besides, this assumption is somewhat questionable given the existence of an ultraviolet (UV) cut-off scale in the effective field theory (EFT) of inflation, beyond which a UV-complete description of quantum gravity is required to properly understand the trans-Planckian dynamics of the modes. Such a non-Bunch-Davies (NBD) framework leads to the choice of instantaneous Minkowski vacua which mix the creation and annihilation operators, thereby rendering each mode function a Bogolyubov rotation of the corresponding BD modes \cite{Danielsson:2002kx}. The Bogolyubov coefficients thus essentially parametrize our ``ignorance'' of the exact dynamics beyond some finite initial past $\eta_0$ where inflation begins, for which there exists a variety of proposed theoretical contenders \cite{Sriramkumar:2004pj,Brandenberger:2009jq,Shiu:2011qw,Dey:2012qp,Sugimura:2013cra,Agullo:2015aba,Choudhury:2017glj}. Such generic initial vacua may lead to interesting features in various primordial correlators, which can either put some theoretical bounds on possible parameter spaces or potentially source a wide range of cosmological observables that could be of interest to future cosmological missions \cite{Chen:2006nt,Holman:2007na,Meerburg,Meerburg:2009fi,Ganc:2011dy,Agullo:2010ws,LopezNacir:2011kk,Brahma:2013rua,Shukla:2016bnu,Chandra:2016jll,Akama:2020jko,Naskar:2020vkd,tahara}. 

The aforementioned considerations motivate us to focus in this work on the three-point cross-correlation function between the primordial curvature perturbation and  the PMF generated via the direct gauge-inflaton coupling scheme, for the case of generic initial vacua (that includes BD as a special case) in both the scalar and the gauge field sectors. One expects the information of pre-inflationary dynamics, that is typically encoded in the extended  parameter space of generic initial vacua, to be translated into detectable cosmological features across subsequent epochs, thereby providing us with valuable windows into primordial physics and the genesis of magnetic fields. Given the expected enhanced sensitivities of next-generation missions which aim to probe a variety of finer cosmological signals and significantly improve the knowledge base of our cosmic history, the origin and signatures of such a correlator merit investigation from both theoretical and observational viewpoints, which underpins the current study as key motivating factors. The present work aims to analyze such a three-point primordial cross-correlation function for all possible triangular configurations with generic Bogolyubov coefficients, and to subsequently explore the potentially observable effects of generic vacua on cosmological observables. As a representative example, in the present article we investigate the non-trivial effects of the generic initial sectors on a possible $C_\ell^{\mu T}$ signal sourced by such a primordial cross-correlation in the presence of PMFs. We further find out possible bounds on the parameters from latest CMB data and, subsequently, search for any possible enhancement in SNR at a few upcoming space-based CMB missions complemented by a forecast study into the chances of detection of such a signal. As such, the current work may be deemed relevant in the era of precision cosmology from multiple perspectives, which essentially add up to the possibility of obtaining an important handle on the physics of PMFs and of non-standard initial vacua, via observable CMB spectral distortion signatures in near future. We believe there is enough room for exploring some other potentially interesting and observationally relevant scenarios for this extended parameter space of generic initial vacua in some of the future cosmological missions. Exciting avenues in view of spectral distortion signatures in particular, which extend beyond the scope of the present work, include studying the prospects of detection at ground-based Stage IV CMB observatories (CMB-S4) \cite{CMB-S4:2016ple,2017arXiv170602464A,Abazajian:2019eic,CMB-S4:2022ght} and at the Square Kilometre Array \cite{Carilli:2004nx,SKA:2018ckk,Weltman:2018zrl,2019arXiv191212699B}, with the latter recently shown to be promising as a CMB experiment in addition to its primary 21-cm science objectives \cite{Zegeye:2024jdc}. We plan to take them up, one at a time, in the upcoming works. 

The paper is organized as follows. In section \ref{sec:recap}, we briefly review the dynamics of the primordial curvature perturbation ($\zeta$) and the gauge field ($A^\mu$) in the direct gauge-inflaton coupled model, by assuming generic initial vacua for both sectors. In section \ref{sec:three_point_corr}, we explicitly compute the analytical form of the NBD three-point correlator $\langle\zeta BB\rangle$ which bears the signature of PMFs, and study its behavior in the different triangular limits, i.e. in the squeezed, equilateral, orthogonal, and flattened configurations. Thereafter in sections \ref{sec:observable_sigs} and \ref{sec:fisher}, we focus on the possible impact of the parameter space corresponding to generic initial conditions on the resulting $C_\ell^{\mu T}$ signal. Our analysis reveals typical enhancement of the SNR up to $\mathcal{O}(10)$ compared to the BD case, corresponding to suitable regions of the NBD parameter space, at several of these missions. We also briefly discuss a few other possible sources of an identical signal and their chances of competing with the aforementioned $C_\ell^{\mu T}$ signal in presence of generic initial vacua. We conclude by summarizing our key results and highlighting a few future directions in section \ref{sec:disc}.

\section{Dynamics of the scalar mode and gauge field} \label{sec:recap}

Let us begin with a brief review of the dynamics of the primordial curvature perturbation ($\zeta$) and the primordial gauge field ($A^\mu$) on the basis of their respective second order actions. This will help us develop the subsequent sections coherently.

\subsection{Scalar curvature mode} \label{subsec:recap_scalar_dynamics}

In the standard inflationary paradigm, the quantum fluctuations of the inflaton field  provide the seeds of perturbation, a convenient description of which is provided by the ADM formalism. Here the quadratic action for the scalar curvature mode ($\zeta$) on an FRLW background with scale factor $a(t)$ can be extracted as 
\begin{equation}
    S_{\zeta\zeta}=\int d^3xd\eta\:a^2\epsilon\:\left(\zeta'^2-\left(\partial_i\zeta\right)^2\right)\:,
\end{equation}
where $\epsilon=-\dot{H}/H^2$ is the first slow-roll parameter, and prime denotes derivative with respect to the conformal time $\eta$. Treating $\zeta(\vec{x},\eta)$ as a quantum field allows a mode expansion of the form
\begin{equation}
    \zeta(\vec{x},\eta)=\int\dfrac{d^3k}{(2\pi)^3}\left[\zeta_k(\eta)e^{-i\vec{k}.\vec{x}}\hat{a}_{\vec{k}}+\zeta^*_k(\eta)e^{i\vec{k}.\vec{x}}\hat{a}^\dagger_{\vec{k}}\right]\:,
\end{equation}
where $[\hat{a}_{\vec{k}_1},\hat{a}^\dagger_{\vec{k}_2}]=(2\pi)^3\delta^{(3)}(\vec{k}_1-\vec{k}_2)$. Defining the canonically conjugate momentum $\Pi^{\scriptscriptstyle (\zeta)}(\vec{x},\eta)=2a^2\epsilon\zeta'(\vec{x},\eta)$, the equal-time commutator $[\zeta(\vec{x},\eta),\Pi(\vec{y},\eta)]=i\delta^{(3)}(\vec{x}-\vec{y})$ leads to a Wronskian condition which serves as an important normalization criterion for the mode function. Extremizing $S_{\zeta\zeta}$, the equation of motion is obtained to be
\begin{equation} \label{scaleqmot}
    \zeta_k''+2aH\left(1+\delta+\epsilon\right)\zeta_k'+k^2\zeta_k=0\:,
\end{equation}
where $\delta=\ddot{H}/{2H\dot{H}}$ is the second slow-roll parameter. Assuming standard BD initial conditions, the normalized solution to \eqref{scaleqmot}, up to the leading order in $\epsilon$ in a quasi-de Sitter background, is given by
\begin{eqnarray}
    \zeta_k^{\scriptscriptstyle \rm (BD)}(\eta)=&&ie^{\frac{i\nu\pi}{2}+\frac{i\pi}{4}}(1+\epsilon)^{\frac{1}{2}-\nu}\left(\dfrac{H}{M_{\textrm{Pl}}}\right) \nonumber \\
    && \times\sqrt{\dfrac{\pi}{8\epsilon}}(-\eta)^\nu H_\nu^{(1)}(-k\eta)\:,
\end{eqnarray}
where $\nu=3/2+\epsilon/(1-\epsilon)$ and $H_\nu^{(1)}$ is the Hankel function of the first kind. For generic initial vacua, however, the solution is modified to $\zeta_k(\eta)=\alpha_k\zeta_k^{\scriptscriptstyle \rm (BD)}(\eta)+\beta_k\zeta_k^{\scriptscriptstyle \rm (BD)*}(\eta)$, where $\alpha_k$ and $\beta_k$ constitute a pair of momentum-dependent complex Bogolyubov coefficients with $|\alpha_k|^2-|\beta_k|^2=1$. The scalar power spectrum on superhorizon scales $k\ll aH$, constructed from the two point correlator of the curvature perturbation, is 
\begin{equation} \label{zetaspec}
    P_\zeta(k)=\dfrac{\Gamma(\nu)^2(1+\epsilon)^{1-2\nu}}{\pi\epsilon k^3}|\alpha_k+\beta_k|^2\left(\dfrac{H}{M_{\rm Pl}}\right)^2\left(\dfrac{k}{2aH}\right)^{n_s-1}\:,
\end{equation}
from which the dimensionless power spectrum follows as $\Delta_s^2(k)=\frac{k^3}{2\pi^2}P_\zeta(k)$, with $n_s=4-2\nu$ being the scalar spectral index. The spectrum is thus predicted to be scale invariant for $\nu=3/2$ in case of a perfectly de Sitter background. However, such a scenario has been ruled out by the latest CMB data from Planck 2018, which has helped narrow down on a tiny but unambiguous scale dependence with $n_s=0.9649\pm0.0042$ (on the basis of the CMB TT+TE+TE+low E+lensing dataset) \cite{pl18_cosm}.
\\

\subsection{Primordial gauge field} \label{subsec:recap_gauge_dynamics}
We consider a direct coupling between the inflaton and the kinetic term of the $U(1)$ gauge field, which breaks the conformal invariance of the Maxwell action and allows amplification of the EM vacuum fluctuations during inflation \cite{ratra}. In the Coulomb gauge $A_0=\partial_iA_i=0$, the EM action can be written as
\begin{eqnarray} \label{S_A}
    S_{A}&&=-\dfrac{1}{4}\int d^4x\sqrt{-g}\:\lambda(\phi)F_{\mu\nu}F^{\mu\nu} \nonumber\\
    &&=\dfrac{1}{2}\int d^3xd\eta\:\lambda(\phi)\left(A_i'^2-\dfrac{1}{2}(\partial_iA_j-\partial_jA_i)^2\right)\:,
\end{eqnarray}
where the coupling function is parametrized as $\lambda(\phi(\eta))=\lambda_I(\eta_I/\eta)^{2n}$, with $\lambda_I\sim1$ to restore classical electrodynamics at the end of inflation. The mode expansion of the gauge field can be written as
\begin{widetext}
\begin{equation}
    A_i(\vec{x},\eta)=\int\dfrac{d^3k}{(2\pi)^3}\sum_s\left[\varepsilon_i^{(s)}(\vec{k})A_k(\eta)e^{-i\vec{k}.\vec{x}}\hat{c}_{\vec{k},s}+\varepsilon_i^{(s)*}(\vec{k})A_k^*(\eta)e^{i\vec{k}.\vec{x}}\hat{c}_{\vec{k},s}^\dagger\right]\:,
\end{equation}
\end{widetext}
where the canonical commutator is $[\hat{c}_{\vec{k},s},\hat{c}_{\vec{k'},s'}^\dagger]=(2\pi)^3\delta_{ss'}\delta^{(3)}(\vec{k}-\vec{k'})$ and $s=\{+,-\}$ are the two polarization indices. The polarization vector satisfies the three conditions of transversality, orthogonality, and completeness. With the conjugate momentum given by $\Pi_j(\vec{x},\eta)=\lambda(\eta)A_j'(\vec{x},\eta)$, the canonical commutation relation $[A_i(\vec{x},\eta),\Pi_j(\vec{y},\eta)]=i\delta_{ij}\delta^{(3)}(\vec{x}-\vec{y})$ provides the Wronskian condition for normalizing $A_k(\eta)$, whose equation of motion is given by
\begin{equation}
    A_k''-\dfrac{2n}{\eta}A_k'+k^2A_k=0\:.
\end{equation}
This admits the normalized BD solution $A_k^{\scriptscriptstyle \rm (BD)}(\eta)=A_{k^*}u_{n+\frac{1}{2}}(-k\eta)$, which is defined explicitly as
\begin{widetext}
\begin{equation} \label{amodesuperhor}
    A_{k^*}=-i\left(\dfrac{2^{n-\frac{1}{2}}\Gamma(n+\frac{1}{2})}{\sqrt{\pi\lambda_I}}\right)\dfrac{(-k\eta_I)^{-n}}{\sqrt{k}}e^{i\pi(n+1)/2}\:\:,\:\:u_a(x)=\dfrac{i\pi x^a}{2^a\Gamma(a)}H_a^{(1)}(x)\:.
\end{equation}
\end{widetext}
In the limit $k\ll aH$, the quantity $A_{k^*}$ plays the role of the constant superhorizon value of $A_k(\eta)$. In case of generic vacua, the complete solution is modified to $A_k(\eta)=\gamma_kA_k^{\scriptscriptstyle \rm (BD)}(\eta)+\delta_kA_k^{\scriptscriptstyle \rm (BD)*}(\eta)$ with $|\gamma_k|^2-|\delta_k|^2=1$. The power spectrum of the gauge field can then be defined as
\begin{eqnarray}
    &&\langle0|A_i(\vec{k},\eta)A_j(\vec{k'},\eta)|0\rangle \nonumber\\
    &&=(2\pi)^3\delta^{(3)}(\vec{k}+\vec{k'})\left(\delta_{ij}-\dfrac{k_ik_j}{k^2}\right)P_A(k,\eta)\:,
\end{eqnarray}
which on superhorizon scales reduces to $P_A(k)=|A_{k^*}|^2|\gamma_k+\delta_k|^2$. Using the definition of the associated magnetic field $B_i(\vec{x},\eta)=a(\eta)^{-1}\varepsilon_{ijk}\partial_jA_k(\vec{x},\eta)$, the magnetic power spectrum can subsequently be derived as
\begin{eqnarray} \label{magspec}
    &&\langle0|B_i(\vec{k},\eta)B^i(\vec{k'},\eta)|0\rangle=(2\pi)^3\delta^{(3)}(\vec{k}+\vec{k'})P_B(k,\eta) \nonumber\\
    &&\implies P_B(k)=\dfrac{2k^2}{a(\eta_I)^4}|A_{k^*}|^2|\gamma_k+\delta_k|^2\:.
\end{eqnarray}
For $n=2$, a scale invariant magnetic power spectrum is produced, which, after redshifting down to the present time, can lead to the required (sub)-nG field strength coherent on Mpc scales without large back reaction of the EM energy density. However, in this picture, the classical electric charge at the beginning of inflation needs to be $q_{in}\gg1$ to settle down later at $q_I\sim1$ around reheating, thereby threatening the perturbative framework of inflationary magnetogenesis. Among the extended scenarios aiming to resolve this strong coupling problem \cite{sawtooth,bounce_fields,resolve_3}, one particular proposal, for instance, appeals to explicitly broken gauge symmetry of the EM action at the energy scales of relevance \cite{Caldwell_Motta,Domenech:2015zzi,Domenech:2017caf}. Such a scenario may emerge from fundamental UV-complete descriptions, whose detailed discussion falls beyond our current scope (an example can be found in \cite{Chu:1994kc,Dubovsky:2001gnx}). In this work, however, our primary focus lies on the analysis of three-point cross-correlations between $\zeta$ and $\vec{B}$, both generated perturbatively and amplified during inflation, under the assumption of generic initial vacua. For the purpose of this study, we assume the existence of such a UV-complete mechanism, which can cure the simplest inflaton-coupling model of the strong coupling pathology at very early epochs without spoiling our key results.

\section{The scalar-magnetic three-point cross-correlator} \label{sec:three_point_corr}

Having set the stage, let us now engage ourselves in explicitly computing the three-point mixed correlator $\langle\zeta BB\rangle$ for generic vacuum (BD/NBD) by employing the ``in-in'' formalism \cite{Maldacena_PNG}. In this picture, the initial vacuum states are evolved up to time $t$ using the interaction Hamiltonian $H_{\textrm{int}}(t)$, and the expectation value of an observable $O(t)$ is taken as 
\begin{widetext}
\begin{equation}
    \langle\Omega| O(t)|\Omega\rangle=\langle 0|\bar{\mathcal{T}}\left(e^{i\int_0^t H_{\textrm{int}}(t_1)dt_1}\right)O(t)\:\mathcal{T}\left(e^{-i\int_0^t H_{\textrm{int}}(t_1)dt_1}\right)|0\rangle\:,
\end{equation}
\end{widetext}
where $|0\rangle$ is the asymptotic vacuum of the free theory, $|\Omega\rangle$ is the instantaneous vacuum of the interacting theory, and $\mathcal{T}$ and  $\bar{\mathcal{T}}$ are the time ordering and anti-time ordering operators respectively. 
Expanding this master formula up to the first order, the leading contribution to our required three-point correlator at the end of inflation ($\eta_I\to0$) is given as
\begin{widetext}
\begin{equation} \label{mastereq}
    \langle \zeta(\vec{k_1},\eta_I)A_i(\vec{k_2},\eta_I)A_j(\vec{k_3},\eta_I)\rangle
    =-2\:\textrm{Im}\int_{\eta_0}^{\eta_I}d\eta_1a(\eta_1)\langle0|H_{\zeta AA}(\eta_1)\zeta(\vec{k_1},\eta_I)A_i(\vec{k_2},\eta_I)A_j(\vec{k_3},\eta_I)|0\rangle\:,
\end{equation}
\end{widetext}
where $\eta_0\to-\infty$ denotes the conformal time at the beginning of inflation. Making use of the definition of the magnetic field, the $\langle\zeta BB\rangle$ correlator can be obtained immediately as
\begin{widetext}
\begin{equation} \label{eq:xbb_contr}
    \langle \zeta(\vec{k_1},\eta_I)B_i(\vec{k_2},\eta_I)B^i(\vec{k_3},\eta_I)\rangle
    =-\dfrac{1}{a(\eta_I)^4}\left(\delta_{ij}\:\vec{k_2}.\vec{k_3}-k_{2j}k_{3i}\right)\langle \zeta(\vec{k_1},\eta_I)A_i(\vec{k_2},\eta_I)A_j(\vec{k_3},\eta_I)\rangle\:.
\end{equation}
\end{widetext}
In what follows we shall make use of this cross-correlator in calculating the bispectra for different triangular configurations of observational interest.

\subsection{The cubic interaction Hamiltonian} \label{subsec:int_hamiltonian}
The most straightforward method of obtaining the third-order interaction action $S_{\zeta AA}$ is to simply expand $S_A$ from \eqref{S_A} up to first order in the metric perturbation as
\begin{eqnarray}
    S_{\zeta AA}=-\dfrac{1}{4}\int &&d^4x\lambda(\eta)\bigg[(\sqrt{-g})^{(1)}(g^{\alpha\mu}g^{\beta\nu})^{(0)} \nonumber\\
    &&+(\sqrt{-g})^{(0)}(g^{\alpha\mu}g^{\beta\nu})^{(1)}\bigg]F_{\alpha\beta}F_{\mu\nu}\:.
\end{eqnarray}
In terms of the ADM decomposed inflationary metric outlined in \ref{subsec:recap_scalar_dynamics}, one then arrives at the interaction Hamiltonian
\begin{eqnarray} \label{intapp}
    H_{\zeta AA}(t)=-\dfrac{1}{2}\int d^3x\:\dfrac{\lambda(\eta)}{a(\eta)}\bigg[&&\left(\zeta-\dfrac{\zeta'}{aH}\right)\left(A_i'^2+\dfrac{1}{2}F_{ij}^2\right) \nonumber\\
    &&+\dfrac{2(\partial_i\zeta)}{aH}A_j'F_{ij}\bigg]\:,
\end{eqnarray}
where $S_{\zeta AA}=-\int dtH_{\zeta AA}(t)$. While this matches the expression derived and used in \cite{Caldwell_Motta}, the true interaction between the curvature perturbation and the gauge field is of lower order, since this action is proportional to a total time derivative at leading order as shown in \cite{Jain_Sloth}. Extracting that out via a field redefinition, the true third-order interaction Hamiltonian turns out to be 
\begin{equation} \label{scalintham}
    H_{\zeta AA}(t)=\dfrac{1}{2}\int d^3x\:\dfrac{\lambda'(\eta)\zeta}{a(\eta)^2H}\left(A_i'^2-\dfrac{1}{2}F_{ij}^2\right)\:,
\end{equation}
where we have made use of the equation of continuity $\nabla_\mu T^{\mu\nu}+\frac{1}{4}(\nabla^\nu\lambda)F_{\rho\sigma}F^{\rho\sigma}=0$.

\subsection{Three-point function for generic initial vacua} \label{subsec:three_point_calc}

Plugging the Hamiltonian \eqref{scalintham} in  \eqref{mastereq}, followed by some straightforward algebra, the $\langle\zeta AA\rangle$ correlator turns out to be
\begin{eqnarray} \label{raacorr}
&&\langle\zeta(\vec{k_1},\eta_I)A_i(\vec{k_2},\eta_I)A_j(\vec{k_3},\eta_I)\rangle=-(2\pi)^3\delta^{(3)}(\vec{k_1}+\vec{k_2}+\vec{k_3}) \nonumber\\
&&\times\bigg[\left(\delta_{il}-\dfrac{k_{2i}k_{2l}}{k_2^2}\right)\left(\delta_{jl}-\dfrac{k_{3j}k_{3l}}{k_3^2}\right)(\mathcal{I}_1+\vec{k_2}.\vec{k_3}\mathcal{I}_2) \nonumber\\
&&-\left(\delta_{il}-\dfrac{k_{2i}k_{2l}}{k_2^2}\right)\left(\delta_{jm}-\dfrac{k_{3j}k_{3m}}{k_3^2}\right)k_{2m}k_{3l}\mathcal{I}_2\bigg]\:,
\end{eqnarray}
where the momentum-conserving delta function results in triangular configurations in $k$-space, and $\mathcal{I}_1$ and $\mathcal{I}_2$ are time integrals emerging from the in-in prescription as
\begin{subequations}
\begin{eqnarray} \label{I_1}
    \mathcal{I}_1&&=2\:\textrm{Im}\bigg[\zeta_{k_1}(\eta_I)A_{k_2}(\eta_I)A_{k_3}(\eta_I) \nonumber\\
    &&\times\int_{\eta_0}^{\eta_I}d\eta_1\eta_1\lambda'(\eta_1)\zeta_{k_1}^*(\eta_1)A_{k_2}^{*'}(\eta_1)A_{k_3}^{*'}(\eta_I)\bigg]\:,
\end{eqnarray}
\begin{eqnarray} \label{I_2}
    \mathcal{I}_2&&=2\:\textrm{Im}\bigg[\zeta_{k_1}(\eta_I)A_{k_2}(\eta_I)A_{k_3}(\eta_I) \nonumber\\
    &&\times\int_{\eta_0}^{\eta_I}d\eta_1\eta_1\lambda'(\eta_1)\zeta_{k_1}^*(\eta_1)A_{k_2}^*(\eta_1)A_{k_3}^*(\eta_I)\bigg]\:.
\end{eqnarray}
\end{subequations}
Consequently, the $\langle\zeta BB\rangle$ correlator can be arrived at by combining \eqref{eq:xbb_contr} and \eqref{raacorr}:  
\begin{eqnarray} \label{rbb}
    \langle &&\zeta(\vec{k_1},\eta_I)B_i(\vec{k_2},\eta_I)B^i(\vec{k_3},\eta_I)\rangle \nonumber\\
    &&=(2\pi)^3 \delta^{(3)}(\vec{k_1}+\vec{k_2}+\vec{k_3})\times\mathcal{B}(k_1,k_2,k_3)\:,
\end{eqnarray}
where the primordial cross-bispectrum $\mathcal{B}(k_1,k_2,k_3)$ has the structure
\begin{eqnarray} \label{rbbbispec}
    &&\mathcal{B}(k_1,k_2,k_3) \nonumber\\
    &&=a(\eta_I)^{-4}\left[2(\vec{k_2}.\vec{k_3})\mathcal{I}_1+\left((\vec{k_2}.\vec{k_3})^2+k_2^2k_3^2\right)\mathcal{I}_2\right]\:.
\end{eqnarray}

It is now time to plug in the explicit solutions for the generic mode functions. 
As a leading order approximation, we neglect the slow-roll parameter in the order of the Hankel functions for the scalar mode, and consider $n=2$ in the gauge field sector which corresponds to the case of a scale-invariant magnetic power spectrum without considerable back reaction. This results in no significant loss of information, and allows us to focus on a situation which is of physical interest. Using the chosen parametrization of the coupling function $\lambda(\eta)$ from \eqref{amodesuperhor} and following some algebraic manipulations, the two time integrals can be written succinctly as
\begin{subequations} \label{I-J}
\begin{equation} 
    \mathcal{I}_1=(+2n)|\zeta_{k_1^*}|^2A_{k_2^*}A_{k_3^*}k_2k_3\mathcal{J}_1\:,
\end{equation}
\begin{equation}
    \mathcal{I}_2=(+2n)|\zeta_{k_1^*}|^2A_{k_2^*}A_{k_3^*}\mathcal{J}_2\:,
\end{equation}
\end{subequations}
where $\zeta_{k^*}\approx\dfrac{1}{\sqrt{4\epsilon k^3}}\left(\dfrac{H}{M_{\rm Pl}}\right)$, $A_{k^*}$ is the superhorizon part of the gauge field from \eqref{amodesuperhor}, and the reduced integrals $\mathcal{J}_1$ and $\mathcal{J}_2$ are given by
\begin{widetext}
\begin{subequations}
\begin{eqnarray} \label{j1nbd}
   \mathcal{J}_1=&&\dfrac{\pi^3}{2}\dfrac{2^{-2n-1}}{\Gamma(n+\frac{1}{2})^2}(-k_2\eta_I)^{n+\frac{1}{2}}(-k_3\eta_I)^{n+\frac{1}{2}}\times \textrm{Im}\Bigg[\left(\alpha_1(1+ik_1\eta_I)e^{-ik_1\eta_I}+\beta_1(1-ik_1\eta_I)e^{ik_1\eta_I}\right) \nonumber\\
   &&\times\left(\gamma_2H_{n+\frac{1}{2}}^{(1)}(-k_2\eta_I)-\delta_2H_{n+\frac{1}{2}}^{(2)}(-k_2\eta_I)\right)\left(\gamma_3H_{n+\frac{1}{2}}^{(1)}(-k_3\eta_I)-\delta_3H_{n+\frac{1}{2}}^{(2)}(-k_3\eta_I)\right) \nonumber\\
   &&\times\int_{\eta_0}^{\eta_I}d\eta_1\eta_1\left(\alpha_1^*(1-ik_1\eta_1)e^{ik_1\eta_1}+\beta_1^*(1+ik_1\eta_1)e^{-ik_1\eta_1}\right)\left(\gamma_2^*H_{n-\frac{1}{2}}^{(2)}(-k_2\eta_1)-\delta_2^*H_{n-\frac{1}{2}}^{(1)}(-k_2\eta_1)\right) \nonumber\\
   &&\times\left(\gamma_3^*H_{n-\frac{1}{2}}^{(2)}(-k_3\eta_1)-\delta_3^*H_{n-\frac{1}{2}}^{(1)}(-k_3\eta_1)\right)\Bigg]\:,
\end{eqnarray}
\begin{eqnarray} \label{j2nbd}
   \mathcal{J}_2=&&\dfrac{\pi^3}{2}\dfrac{2^{-2n-1}}{\Gamma(n+\frac{1}{2})^2}(-k_2\eta_I)^{n+\frac{1}{2}}(-k_3\eta_I)^{n+\frac{1}{2}}\times\textrm{Im}\Bigg[\left(\alpha_1(1+ik_1\eta_I)e^{-ik_1\eta_I}+\beta_1(1-ik_1\eta_I)e^{ik_1\eta_I}\right)\nonumber\\ &&\times\left(\gamma_2H_{n+\frac{1}{2}}^{(1)}(-k_2\eta_I)-\delta_2H_{n+\frac{1}{2}}^{(2)}(-k_2\eta_I)\right)\left(\gamma_3H_{n+\frac{1}{2}}^{(1)}(-k_3\eta_I)-\delta_3H_{n+\frac{1}{2}}^{(2)}(-k_3\eta_I)\right) \nonumber\\
   &&\times \int_{\eta_0}^{\eta_I}d\eta_1\eta_1\left(\alpha_1^*(1-ik_1\eta_1)e^{ik_1\eta_1}+\beta_1^*(1+ik_1\eta_1)e^{-ik_1\eta_1}\right)\left(\gamma_2^*H_{n+\frac{1}{2}}^{(2)}(-k_2\eta_1)-\delta_2^*H_{n+\frac{1}{2}}^{(1)}(-k_2\eta_1)\right) \nonumber\\
   &&\times\left(\gamma_3^*H_{n+\frac{1}{2}}^{(2)}(-k_3\eta_1)-\delta_3^*H_{n+\frac{1}{2}}^{(1)}(-k_3\eta_1)\right)\Bigg]\:.
\end{eqnarray}
\end{subequations}
\end{widetext}
where we have used the shorthand $\alpha_{k_1}\to\alpha_1$ and so on for all the momentum-labeled Bogolyubov coefficients for notational convenience. Although somewhat cumbersome at first glance, these expressions can be compared directly with \eqref{I_1} and \eqref{I_2}, which shows that we must have a product of the NBD scalar mode function and two powers of the NBD gauge field mode function, both within and outside each of the integrands separately. The pre-factors follow from the superhorizon limits quoted earlier. The full expressions of these integrals for arbitrary triangular configurations, while analytically computable, are quite cumbersome and proffer little physical insight on their own. As such, we have included them in appendix \ref{sec:appendix} for the interested reader.  For now, we move on to the various limiting cases of our interest.

\subsection{Limiting triangular configurations} \label{subsec:lim_triangles}

As is well-known, there are
four major limiting triangular shapes of interest, namely, the squeezed, equilateral, orthogonal, and flattened configurations, which can be used to approximate a wide range of different possible shapes of the bispectrum. We shall focus on the behavior of the $\langle\zeta BB\rangle$ correlator for generic vacua in each of these limits separately. 

\subsubsection*{Squeezed limit}

In the \textit{squeezed} limit, where $k_1\ll k_2\approx k_3$ and $\vec{k_2}\approx-\vec{k_3}$, the expression from \eqref{rbbbispec} reduces to
\begin{eqnarray} \label{sqlimrbb}
    \mathcal{B}^{\rm (sq)}(k_1,k_2,k_3)=\:&&|\zeta_{k_1^*}|^2|A_{k_2^*}|^2\times\dfrac{2n}{a(\eta_I)^4} \nonumber\\
    &&\times 2k_2^4\left(-\mathcal{J}_1^{\rm (sq)}+\mathcal{J}_2^{\rm (sq)}\right)\:.
\end{eqnarray}
Plugging in the squeezed limit approximations $\mathcal{J}_1^{\rm (sq)}$ and $\mathcal{J}_2^{\rm (sq)}$ of the two integrals where we neglect terms of order $\mathcal{O}(k_1/k_2)$ and beyond, we finally obtain
\begin{equation} \label{sqlim}
    \mathcal{B}^{\rm (sq)}(k_1,k_2,k_3)=b_{\rm NL}^{\rm (loc)}P_\zeta(k_1)P_B(k_2)\:,
\end{equation}
where the power spectra for generic initial conditions are defined according to \eqref{zetaspec} and \eqref{magspec}, and $b_{\rm NL}^{\rm (loc)}=2n\:(=4)$ plays the role of a momentum-independent local-type nonlinearity parameter for the mixed correlator in this limit. 

A few comments about this result are in order. While \eqref{sqlim} apparently boils down to the same two-point factorizable form as obtained for the BD case in \cite{Jain_Sloth}, this reduction is actually quite non-trivial as no assumption has been made so far as to the details of the initial vacua, which have instead been kept as generic as possible. While semi-classical arguments put forward in earlier works attempt to shed light on such local-type non-Gaussian behavior in the standard BD scenario \cite{Jain:2012ga,Jain_Sloth,Jain_grav}, it is worth stressing here that such arguments may need to be revised when non-standard initial vacua come into play. For example, similar calculations with the tensor perturbation ($h_{ij}$) for the same background model reveal that the squeezed limit $\langle hBB\rangle$ correlator, which is proportional to $P_h(k_1)P_B(k_2)$ and admits a similar local-type interpretation in the BD case, actually experiences an additional $\mathcal{O}(k_2/k_1)$ enhancement when NBD initial vacua are considered. Semi-classical arguments to explain the local-type behavior of $\langle hBB\rangle$ in the squeezed limit thus falter in the generic case. While we intend to focus on the nature and possible detectable signatures of the NBD $\langle hBB\rangle$ correlator separately in a future work, this particular feature is stated here solely to emphasize that the formal similarity of expressions between the fully generic result given in \eqref{sqlim} and its BD limit from \cite{Jain_Sloth} is not an obvious property of such correlators.

Perhaps the most important comment on \eqref{sqlim} would be based on its prospects of measurements in present and future cosmological missions. It should be noted that even though \eqref{sqlim} looks similar to the one in BD case, as illustrated above, the cross-bispectrum results in  non-trivial effects so far as its observational aspects are concerned, mostly due to the fact that it intrinsically takes into account the effects of generic initial states. This expanded parameter space for the Bogolyubov coefficients (which are not necessarily BD now) helps us explore their observational prospects in the light of upcoming CMB observations. One such possibility will be demonstrated in detail in the following sections. We hope to come up with some more prospects in the near future. 

\subsubsection*{Flattened limit}

In the \textit{flattened} limit $k_1\approx2k_2=2k_3$, both $\mathcal{J}_1$ and $\mathcal{J}_2$ apparently diverge at first glance due to the presence of terms proportional to $(k_2+k_3-k_1)^{-1}=\tilde{k}_1^{-1}$. This, however, is a well-known feature of primordial three-point correlators when generic initial states are involved \cite{Holman:2007na,Chen:2006nt}. The origin of the divergence can be traced to the lower limit of past infinity imposed on the integrals, which, as highlighted earlier, is no longer physically tenable in the NBD scenario but needs to be replaced with a finite past cut-off $\eta_0$. While $\eta_0\to-\infty$ is computationally valid for the other triangular shapes, a finite $\eta_0$ highly suppresses the oscillatory nature of the $\exp({\pm i\tilde{k}_1\eta_0})$ factors in the integrals in the limit $\tilde{k}_1\to0$, thereby enhancing the contribution of such terms to the bispectrum. This in turn acts as a regulating mechanism to cure the apparent divergence in the flattened limit, resulting instead in $\mathcal{O}(|k\eta_0|^p)$ enhancements, where the exponent $p$ arises from the specific structure of the correlator. 

Assuming weak deviation from BD initial vacua, i.e. $\beta_i,\delta_j\ll1$, one can notice similar behavior shown by the flattened-limit $\langle\zeta BB\rangle$ correlator. Expanding $\mathcal{J}_1^{\rm (fl)}$ and $\mathcal{J}_2^{\rm (fl)}$ up to the first order in the NBD coefficients, the relevant terms which carry $\tilde{k}_1$ in the denominator in both cases are the integrals proportional to $\beta_1^*$, given by
\begin{widetext}
\begin{subequations}
\begin{equation} \label{flatint1}
    \begin{split}
        &\dfrac{\pi}{2}\textrm{Im}\int_{\eta_0}^{\eta_I}d\eta_1\eta_1(1+ik_1\eta_1)e^{-ik_1\eta_1}H_\frac{3}{2}^{(2)}(-k_2\eta_1)H_\frac{3}{2}^{(2)}(-k_3\eta_1) \\
        & =\dfrac{1}{(k_2k_3)^\frac{3}{2}}\left[-\dfrac{\mathcal{K}_1}{\tilde{k}_1^2}\left(1-\cos(\tilde{k}_1\eta_0)\right)+\tilde{k}_1\left(1-\dfrac{\sin(\tilde{k}_1\eta_0)}{\tilde{k}_1\eta_0}\right)+\dfrac{\sin(\tilde{k}_1\eta_0)}{\tilde{k}_1}(k_1k_2k_3)\eta_0\right]\:,
    \end{split}
\end{equation}
\begin{equation} \label{flatint2}
    \begin{split}
        &\dfrac{\pi}{2}\textrm{Im}\int_{\eta_0}^{\eta_I}d\eta_1\eta_1(1+ik_1\eta_1)e^{-ik_1\eta_1}H_\frac{5}{2}^{(2)}(-k_2\eta_1)H_\frac{5}{2}^{(2)}(-k_3\eta_1) \\
        & =\dfrac{1}{(k_2k_3)^\frac{5}{2}}\left[-\dfrac{k_2k_3\mathcal{K}_2}{\tilde{k}_1^2}\left(1-\cos(\tilde{k}_1\eta_0)\right)+\dfrac{\sin(\tilde{k}_1\eta_0)}{\tilde{k}_1}(k_1k_2^2k_3^2)\eta_0-k_1^3\ln\left(\dfrac{\eta_I}{\eta_0}\right)\right]\:,
    \end{split}
\end{equation}
\end{subequations}
\end{widetext}
where $\mathcal{K}_1=k_1^2(k_2+k_3)+k_2^2(k_3-k_1)+k_3^2(k_2-k_1)-4k_1k_2k_3$ and $\mathcal{K}_2=3k_1^2(k_2+k_3)+k_2k_3(k_2+k_3)-k_1(3k_2^2+8k_2k_3+3k_3^2)$, and $\ln(\eta_I/\eta_0)=-N_{I}$ is the number of $e$-folds of inflation. The last term arises from the asymptotic behavior of the non-vanishing exponential integral function $\textrm{Ei}(-i\tilde{k}_1\eta_1)$ in the full expression of $\mathcal{J}_2$. In the limit $\tilde{k}_1\to0$, it is easy to see that both \eqref{flatint1} and \eqref{flatint2} are dominated by $\eta_0^2$. Adding the two contributions according to \eqref{rbbbispec}, the leading order NBD correction to the flattened limit bispectrum for $n=2$ is obtained as
\begin{eqnarray}
    \mathcal{B}^{\rm (fl)}(k_1,k_2,k_3)&&\:\approx\mathcal{B}^{\rm (fl)}_{BD}(k_1,k_2,k_3)\nonumber\\
    &&+\dfrac{4}{a(\eta_I)^4}|\zeta_{k_1^*}|^2|A_{k_2^*}|^2\beta_1^*k_1^2(k_1\eta_0)^2\:,
\end{eqnarray}
where $\mathcal{B}^{\rm (fl)}_{BD}(k_1,k_2,k_3)$ is the standard BD value of the correlator. Thus, the enhancement factor due to non-standard initial vacua comes with the exponent $p=2$ in the flattened limit, which should further reduce to $p=1$ while taking the angular average for CMB observables \cite{Holman:2007na}. Possible signatures of such an NBD correction on CMB features could thus be an interesting area to explore in the future. 

The scalar-magnetic cross-bispectrum can also be evaluated in a straightforward manner for the equilateral configuration ($k_1=k_2=k_3$) and for the orthogonal configuration ($k_1=\sqrt{2}k_2=\sqrt{2}k_3$). But the final results for these two configurations do not turn out to be as immediately transparent as the squeezed and flattened cases. Nevertheless, they exhibit some interesting features. As such, we have included the results of these two cases in Appendix \ref{sec:appendix_2}.

\section{Observable imprint on CMB \texorpdfstring{$\mu T$}{} correlation} \label{sec:observable_sigs}

We are now in a position to demonstrate how the presence of generic initial vacua in case of the $\langle\zeta BB\rangle$ correlator can discernibly affect the CMB signal, which might be of interest to current and upcoming experiments. For the purpose of the current work, we focus on the two-point angular cross-spectrum $C_\ell^{\mu T}$, which involves CMB temperature anisotropies sourced by the curvature perturbation, and $\mu$-type spectral distortions sourced by the damping of primordial magnetic fields in the photon-baryon fluid during the pre-recombination era. Overall, such a CMB correlation can be generated by the squeezed limit bispectrum $\mathcal{B}^{\rm (sq)}(k_1,k_2,k_3)$, where two short wavelength magnetic modes remain correlated with a scalar curvature mode of much longer wavelength.

\subsection{Magnetic field-induced cross-power spectrum \texorpdfstring{$C_\ell^{\mu T}$}{}}

In  calculating the $C_\ell^{\mu T}$ function, we closely follow the formalism presented in \cite{ganc_sloth}. The energy density of the comoving magnetic field $\vec{b}(\vec{x},t)=a(t)^2\vec{B}(\vec{x},t)$ is given by
\begin{eqnarray}
    &&\rho_B^{(co)}(\vec{x},t)=\dfrac{\vec{b}^*(\vec{x},t).\vec{b}^*(\vec{x},t)}{2\mu_0} \nonumber\\
    &&=\dfrac{1}{2\mu_0}\int\dfrac{d^3k_3d^3k_2}{(2\pi)^6}\vec{b}^*(\vec{k}_3).\vec{b}^*(\vec{k}_2)e^{-\frac{k_3^2+k_2^2}{k_D(t)^2}}e^{-i(\vec{k}_3+\vec{k}_2).\vec{x}}\:,
\end{eqnarray}
where $\mu_0$ is the magnetic permeability of vacuum, and $k_D(t)$ is the magnetic damping scale over the $\mu$-distortion epoch. As this dissipated energy ($\Delta E$) is injected into the photon-baryon plasma, the CMB spectrum effectively shifts from a near-perfect Planckian spectrum to a Bose-Einstein spectrum, with a non-zero chemical potential given by $\mu=1.4\times(\Delta E/E)$. Making use of the plane wave expansion formula, the $\mu$-multipoles are subsequently given by
\begin{eqnarray} \label{almmu}
    a_{\ell m}^\mu=&&1.4\times(-1)^{\ell+m}\dfrac{4\pi i^\ell}{2\mu_0\rho_{\gamma0}}\int\dfrac{d^3k_3d^3k_2}{(2\pi)^6}\vec{b}^*(\vec{k}_3).\vec{b}^*(\vec{k}_2) \nonumber\\
    &&\times\left[e^{-\frac{k_3^2+k_2^2}{k_D^2}}\right]_f^iW\left(\dfrac{k_+}{k_s}\right)Y_{\ell m}(\hat{k}_+)j_\ell(k_+r_L)\:,
\end{eqnarray}
where $\vec{k}_+=\vec{k}_3+\vec{k}_2$, $j_\ell(x)$ is the spherical Bessel function of order $\ell$, the window function $W(x)=3x^{-3}(\sin x-x\cos x)$ is introduced for smoothing over the $\mu$-dissipation scale, $\rho_{\gamma0}$ is the photon energy density of the CMB at the present epoch, and $r_L$ is the distance to the surface of last scattering. Typical values of these quantities estimated from available data have been used in the next section.

On the other hand, the temperature anisotropy multipoles are given in terms of the source curvature perturbation as
\begin{equation} \label{almt}
    a_{\ell m}^T=\dfrac{12\pi}{5}i^\ell\int\dfrac{d^3k_1}{(2\pi)^3}\zeta_{\vec{k}_1}\Delta^{T}_{\ell}(k_1)Y_{lm}^*(\hat{k}_1)\:,
\end{equation}
where $\Delta^{T}_{\ell}(k)$ is the scalar transfer function. Defining the cross-power spectrum between temperature anisotropies and $\mu$-type spectral distortions as $\langle a_{\ell m}^{\mu*}a_{\ell'm'}^T\rangle=C_\ell^{\mu T}\delta_{\ell\ell'}\delta_{mm'}$, and recalling the form of the squeezed limit primordial bispectrum $\mathcal{B}^{\rm (sq)}(k_1,k_2,k_3)$ from \eqref{sqlim}, one obtains
\begin{eqnarray} \label{clmut}
    C_\ell^{\mu T}=&&1.4\times\dfrac{12}{5(2\pi)^3}\dfrac{b_{\rm NL}^{\rm (loc)
}}{\mu_0\rho_{\gamma0}}\int\limits_{\tilde{k}_D^f}^{\tilde{k}_D^i}dk_1k_1^2P_B(k_1)\nonumber\\
&&\times \int\limits_0^{10k_s}dkk^2P_\zeta(k)W\left(\dfrac{k}{k_s}\right) j_\ell(kr_L)\Delta^{T}_{\ell}(k)\:.
\end{eqnarray}
In this expression, the following approximations have been implemented. In the first integral, suppression caused by $W(x)$ beyond $k\sim k_s$ has been modeled with an effective upper cut-off around $k\sim10k_s$. Albeit a crude approximation, this suffices for our current purpose of estimating the orders of magnitude of the SNR for certain benchmark values of the model parameters at future missions, and hence commenting on their relative performances. More accurate analyses of the distortion transfer function could be carried out by considering the full numerically solved evolution of the spectral distortion signal with the aid of specialized codes such as CosmoTherm \cite{cosmotherm} and/or SZPack \cite{szpack}, as demonstrated for example by \cite{Kite:2022eye}. We hope to come up with a future work dedicated to a detailed investigation of this.
\footnote{We are thankful to the anonymous referee for bringing these codes to our notice.} On the other hand, in the second integral, $\tilde{k}_D^i=k_D^i/\sqrt{2}$ and $\tilde{k}_D^f=k_D^f/\sqrt{2}$ are damping scales appropriately normalized for the average magnetic energy density, that arise as limits due to the exponential damping factor from \eqref{almmu}. Plugging in the forms of the scalar and magnetic spectra then allows us to explicitly compute the correlation as
\begin{eqnarray} \label{clmut_interm}
    C_\ell^{\mu T}\approx&&\:1.4\left(\dfrac{24\pi}{5}\right)\left(\dfrac{\widetilde{B}_\mu^2}{\mu_0\rho_{\gamma0}}\right)k_sk_p^{1-n_s}A_s \nonumber\\
    &&\times\int\limits_0^{10k_s}dk_1k_1^{n_s-3}j_1\left(\dfrac{k_1}{k_s}\right)j_\ell(kr_L)\Delta^T_\ell(k)\:,
\end{eqnarray}
where $\widetilde{B}_\mu$ is the present-day magnetic field strength on $\mu$-distortion scales, defined as
\begin{eqnarray} \label{bmusq}
    &&\widetilde{B}_\mu^2\equiv\int\limits_{\tilde{k}_D^f}^{\tilde{k}_D^i}\dfrac{d^3k_2}{(2\pi)^3}P_B(k_2)\times\left(\dfrac{a_I}{a_0}\right)^4 \nonumber\\
    &&=\dfrac{9M_{\rm Pl}^4}{2\pi^2\lambda_I}|\gamma_2+\delta_2|^2\left(\dfrac{H}{M_{\rm Pl}}\right)^4\ln\left(\dfrac{\tilde{k}_D^i}{\tilde{k}_D^f}\right)\left(\dfrac{a_I}{a_0}\right)^4\:.
\end{eqnarray}
Here, $(a_I/a_0)$ is the ratio of scale factors at the end of inflation and at the present epoch, which, owing to the conservation of entropy, can be approximated by \cite{Chakraborty:2018dmj}
\begin{equation}
    \dfrac{a_I}{a_0}\approx(0.9\times10^{29})^{-1}\left(\dfrac{H}{10^{-5}M_{\rm Pl}}\right)^{-1/2}\:.
\end{equation} 
The field strength $\widetilde{B}_\mu$ can then be expressed in units of G using the value of $M_{\rm Pl}\approx1.2\times10^{19}$ GeV and the conversion formula $1\:\textrm{G}^2/(8\pi)\approx1.91\times10^{-40}\:\textrm{GeV}^4$. On the other hand, the scalar power amplitude is given by
\begin{equation} \label{A_s}
    A_s=\dfrac{[\Gamma(\nu)]^2(1+\epsilon)^{1-2\nu}}{2^{4-2\nu}\pi^3\epsilon}|\alpha_1+\beta_1|^2\left(\dfrac{H}{M_{\rm Pl}}\right)^2\:.
\end{equation}
In deriving the expressions above, we have assumed a quasi-constant momentum profile of the Bogolyubov coefficients for simplicity \cite{Holman:2007na}. This henceforth makes their momentum labels somewhat superfluous, but we shall retain them nonetheless as an indicator of their original sectors. 

\subsection{Possible bounds from current CMB data}

\begin{figure*}
    \centering
    \begin{subfigure}[b]{0.47\textwidth}
        \centering
        \includegraphics[width=\textwidth]{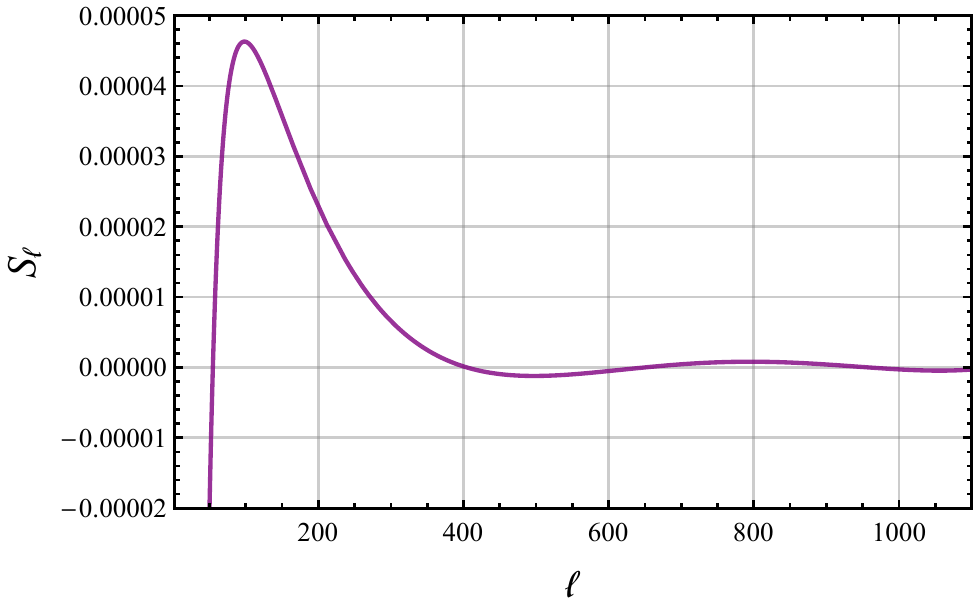}
        \caption{Linearly scaled shape of $S_\ell$.}
    \end{subfigure}%
    ~ 
    \begin{subfigure}[b]{0.46\textwidth}
        \centering
        \includegraphics[width=\textwidth]{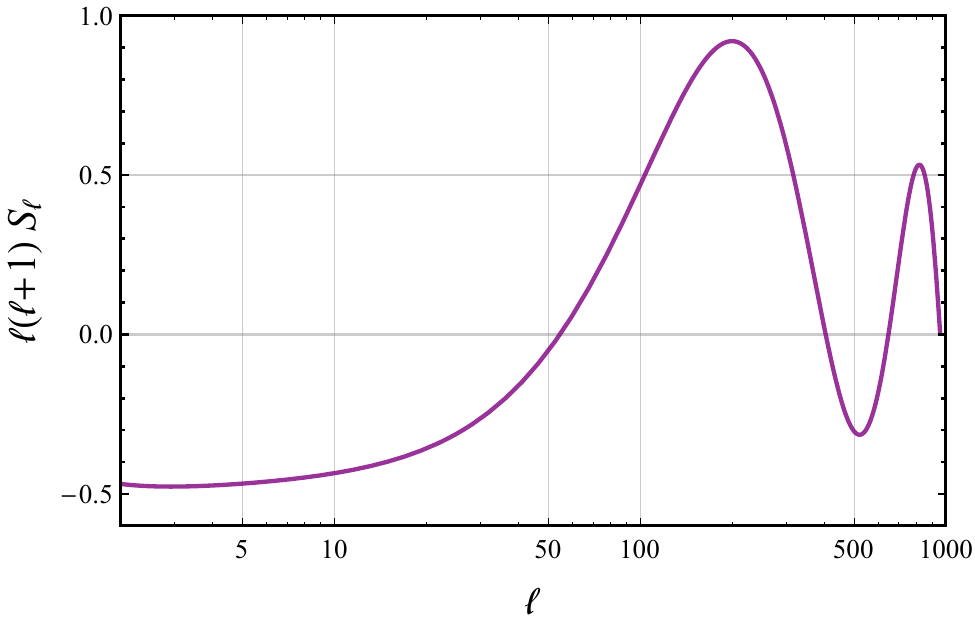}
        \caption{Logarithmically scaled shape of $\ell(\ell+1)S_\ell$.}
    \end{subfigure}
    \caption{Spectral shape of $S_\ell$ appearing in \eqref{clmutfin} as a function of the multipole $\ell$.}
    \label{fig:sl}
\end{figure*}

Let us now estimate the possible bounds on the magnetic field-induced $\mu T$ correlation using latest observational data. The smoothing scale turns out to be approximately $k_s\approx0.084$ Mpc\textsuperscript{-1} following the analysis of \cite{Pajer:2012qep}, while the magnetic damping scale $k_D(t)$ is estimated to be roughly $k_D^i\approx2.1\times10^4$ Mpc\textsuperscript{-1} and $k_D^f\approx83$ Mpc\textsuperscript{-1} at the beginning and the end of the $\mu$-distortion epoch respectively \cite{ganc_sloth}. Based on Planck 2018 TT+TE+EE+low E+lensing data which gives $n_s=0.9649\pm0.0042$ and $A_s=(2.100\pm0.030)\times10^{-9}$ \cite{pl18_inf}, one may constrain the product of the scalar Bogolyubov pre-factor and the Hubble scale of inflation from \eqref{A_s}, which we shall demonstrate shortly. For now, plugging in the mean value of $A_s$, alongside $\rho_{\gamma0}\approx4.1806\times10^{-13}$ erg/cm\textsuperscript{3} and $\mu_0=1$ (in Gaussian units), we arrive at
\begin{eqnarray} \label{clmutfin}
    C_\ell^{\mu T}\approx\:&&\left(1.34\times10^{-7}\right)\times\left[|\gamma_2+\delta_2|\left(\dfrac{H}{M_{\rm Pl}}\right)\right]^2 \nonumber\\
    &&\times\int\limits_0^{10k_s}dk_1k_1^{n_s-3}j_1\left(\dfrac{k_1}{k_s}\right)j_\ell(kr_L)\Delta^T_\ell(k)\:,
\end{eqnarray}
where $\left[|\gamma_2+\delta_2|\left(H/M_{\rm Pl}\right)\right]^2\equiv\theta_{\scriptscriptstyle{B}}$ plays the role of the effective nonlinearity parameter which governs the strength of the correlation. While the Sachs-Wolfe approximated $\Delta^T_\ell(k)\approx-j_\ell(kr_L)/3$ works well for $10\lesssim \ell\lesssim50$, a more accurate numerical solution of the Einstein-Boltzmann equations becomes necessary at higher multipoles. In subsequent calculations, we make use of the full numerically solved $\Delta^T_\ell(k)$, computed over $2\leq \ell\leq1200$ using the publicly available Boltzmann solver code CAMB \cite{camb1,camb2}, for a standard $\Lambda$CDM background cosmology with Planck 2018 TT+TE+EE+low E+lensing best fit parameters (the same also provides us with $r_L\approx13.9$ Gpc). The resulting spectral shape of the integral ($S_l$) is plotted in figure \ref{fig:sl}, which therefore happens to be identical to the shape of $C_\ell^{\mu T}$ apart from an overall normalization.
For $\ell\lesssim1200$,  we find that it can be approximated analytically by the interpolating function
\begin{equation} \label{anapprox}
    S_\ell=\dfrac{\bar{A}}{\ell^3}+\dfrac{\bar{B}}{\ell^2}+\bar{C}+\bar{D}\ell^{\scriptscriptstyle{3/2}}+\bar{E}\ell^4+\bar{F}\ln \ell+\bar{G}Y_2\left(\dfrac{\ell}{1000}\right)\:,
\end{equation}
with the values of the fitting parameters listed in table \ref{tab:fitting}. The polynomial terms, consisting of both positive and negative powers in $\ell$, are essential for the first transition of $S_\ell$ from the negative to the positive domain around $\ell\sim50$. Note that the contributions of the positive powers remain suppressed by the tiny values of $\bar{D}$ and $\bar{E}$ so as to provide a suitable fit. The logarithmic term and the Bessel function term, on the other hand, describe together the subsequent oscillations between the positive and negative domains, with the amplitudes of the peaks and troughs modulated by the values of the coefficients.
\begin{table*}
\caption{\label{tab:fitting}Values of the fitting parameters for the analytical approximation \eqref{anapprox} of $S_\ell$.}
\begin{ruledtabular}
\begin{tabular}{ccccccc}
        {} & {} & {} & {} & {} & {} & {} \\
        $\bar{A}$ & $\bar{B}$ & $\bar{C}$ & $\bar{D}$ & $\bar{E}$ & $\bar{F}$ & $\bar{G}$ \\ 
        {} & {} & {} & {} & {} & {} & {} \\
        \hline
        {} & {} & {} & {} & {} & {} & {} \\
        $0.285105$ & $-8394.82$ & $-0.00152879$ & $-1.08752\times10^{-8}$ & $1.10196\times10^{-16}$ & $-0.000105067$ & $-0.00659292$ \\
        {} & {} & {} & {} & {} & {} & {} \\
\end{tabular}
\end{ruledtabular}
\end{table*}

\begin{figure*}
    \centering
    \begin{subfigure}[b]{0.5\textwidth}
        \centering
        \includegraphics[width=\textwidth]{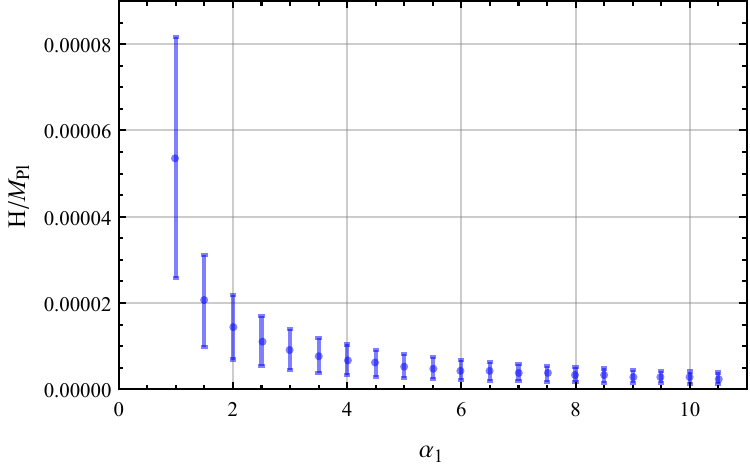}
        \caption{Variation of Hubble parameter during inflation for $\beta_1>0$}
    \end{subfigure}%
    ~ 
    \begin{subfigure}[b]{0.5\textwidth}
        \centering
        \includegraphics[width=\textwidth]{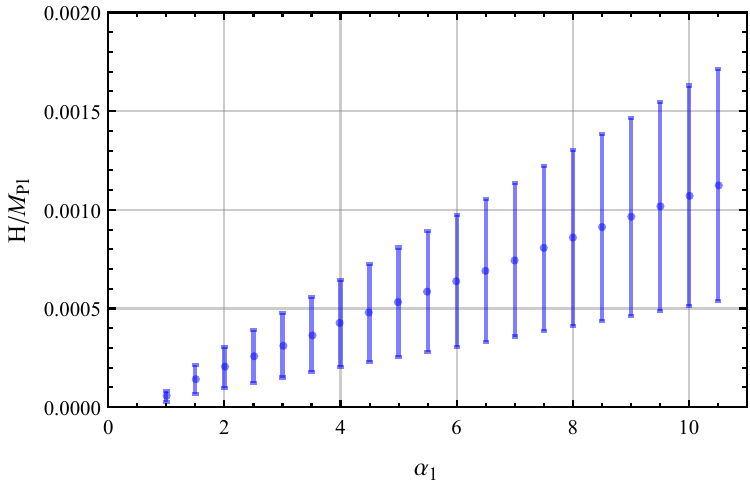}
        \caption{Variation of Hubble parameter during inflation for $\beta_1<0$}
    \end{subfigure}
    \caption{For a range of positive-valued $\alpha_1\in[1,11]$, the derived constraint on $\left(H/M_{\rm Pl}\right)$ consistent with the current bound \eqref{pl18bound_H}, obtained on the basis of Planck 2018 TT+TE+EE+low E+lensing data. Note that the $y$-axis  has been stretched  by a factor of 10 (hence, the error bars are magnified by the same factor) for better visibility. 
    }
    \label{fig:hubble}
\end{figure*}

The $C_\ell^{\mu T}$ signal is thus directly sensitive to the magnetic Bogolyubov parameters through the dependence on $\theta_{\scriptscriptstyle{B}}$ in \eqref{clmutfin}. The role of the scalar initial vacuum is slightly more subtle, and appears through a corresponding tuning of $(H/M_{\rm Pl})$ in order to keep $A_s$ within its observed bounds,  as required by \eqref{A_s}. Using the Planck 2018   constraints on the mean values and $1\sigma$ errors of $A_s$ and $n_s$, we arrive at the relation \footnote{This constraint equation is deduced on the basis of existing bounds on the scalar sector parameters $A_s$ and $n_s$ alone. If one also wishes to take the current upper bound on the tensor-to-scalar ratio ($r$) into account, a possible alternative approach is outlined in appendix \ref{sec:appendix_3}, which effectively produces a  constraint of the same order as in \eqref{pl18bound_H}.}
\begin{equation} \label{pl18bound_H}
    |\alpha_1+\beta_1|\left(\dfrac{H}{M_{\rm Pl}}\right)=(5.37\pm0.28)\times10^{-5}\:.
\end{equation}
Assuming real values of $\alpha_1$ and $\beta_1$, for which the Wronskian condition gives $\beta_1=\pm\sqrt{\alpha_1^2-1}$, we graphically show the resulting constraint on the Hubble scale of inflation due to possible deviations from scalar BD vacuum in figure \ref{fig:hubble}.
Thus, assuming generic initial vacua in both the scalar and the magnetic sectors, one first needs to compute the corresponding mean value of $\left(H/M_{\rm Pl}\right)$ from \eqref{pl18bound_H}, before plugging it in $\theta_{\scriptscriptstyle{B}}$ and evaluating the fiducial $C_\ell^{\mu T}$ according to \eqref{clmutfin}. For a given $\left(H/M_{\rm Pl}\right)$, the choice of the magnetic Bogolyubov parameters is restricted on the other hand by extant COBE/FIRAS+TRIS limits on the sky-averaged monopole $|\mu|<6\times10^{-5}$ \cite{cobe_firas,tris}, which translates to an upper limit of $\widetilde{B}_\mu\lesssim27$ nG (see \eqref{bmusq}). 
\begin{figure*}
    \includegraphics[width=0.9\textwidth]{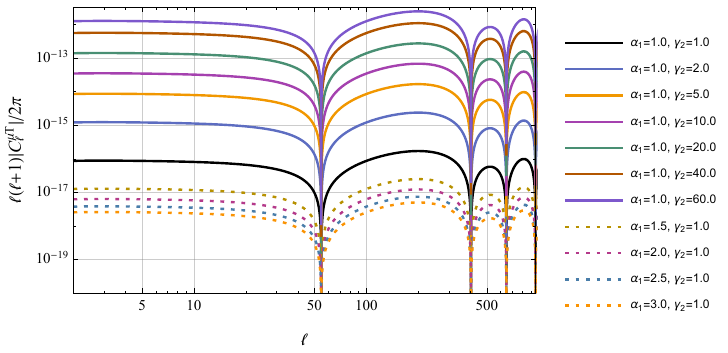}
    \caption{Dependence of the strength of $C_\ell^{\mu T}$ on a few representative positive values of the Bogolyubov coefficients (with $\beta_1=\sqrt{\alpha_1^2-1}$ and $\delta_2=\sqrt{\gamma_2^2-1}$).}
    \label{fig:clmut}
\end{figure*}
In figure \ref{fig:clmut}, we plot the strength of the $C_\ell^{\mu T}$ correlation for a few representative combinations of the Bogolyubov parameters allowed by the aforementioned conditions from COBE/FIRAS+TRIS and Planck 2018 TT+TE+EE+low E+lensing data taken together. The spectral shapes and amplitudes of the $C_\ell^{\mu T}$ signal, obtained by us for the chosen benchmark values of the various model parameters, are largely consistent with the current bounds on $C_\ell^{\mu T}$ obtained by analyzing available sky maps of CMB $\mu$-distortions and temperature anisotropies. For instance, one may refer to  \cite{Khatri:2015tla,Rotti:2022lvy,Bianchini:2022dqh}, whose final results show some noticeable differences owing to CMB deprojection techniques and data analysis pipelines. Nevertheless, the $C_\ell^{\mu T}$ profiles that we obtain in our present work follow the expected spectral shape of the $C_\ell^{\mu T}$ signal and the typical upper bound on its amplitude reported by each of these studies separately. 

A brief discussion on the major outcome of the above figure is in order. From the figure, it transpires that for a BD scalar sector and an NBD gauge field sector, the amplitude of the signal can be significantly enhanced while being consistent with the extant upper bound of $\widetilde{B}_\mu\lesssim27$ nG based on COBE/FIRAS+TRIS data. On the other hand, for an NBD scalar sector and a BD gauge field sector, lowering of $\left(H/M_{\rm Pl}\right)$ associated with the former leads to almost equivalent suppression of the signal. The vanilla BD scenario for both the sectors lies in between.  Another noteworthy point here is that up to the leading order considered here, NBD contributions amount to an overall rescaling of the amplitude of $C_\ell^{\mu T}$ without any phase shift compared to the BD scenario. This can be traced back to the local-type non-Gaussian behavior of $\mathcal{B}^{\rm (sq)}(k_1,k_2,k_3)$ and the assumption of negligible momentum-dependence of the Bogolyubov parameters.

The above analysis with representative values of the Bogolyubov parameters in the light of latest CMB data gives us a clear hint to explore the prospects of the detectability of the scenario in the upcoming CMB missions. We will engage ourselves in that in considerable detail in the following section. 

\section{SNR estimates for future CMB missions} \label{sec:fisher}

In this section, we investigate the prospects of detection of such magnetic field-induced $\mu T$ correlation at a few upcoming CMB missions.
To this end, we will calculate the dependence of the Fisher signal-to-noise ratio (SNR) of the $C_\ell^{\mu T}$ signal on the parameters describing the initial vacua. Given a set of distinct CMB power spectra labeled as $C_\ell^{XX'}$ and a parameter vector $\vec{\theta}$, the Fisher matrix is defined as \cite{Tegmark:1996bz,perotto} 
\begin{equation}
    F_{ij}=\sum_\ell\sum_{PP',QQ'}\dfrac{\partial C_\ell^{PP'}}{\partial\theta_i}\left(\textrm{Cov}_\ell^{-1}\right)_{PP',QQ'}\dfrac{\partial C_\ell^{QQ'}}{\partial\theta_j}\:,
\end{equation}
where $\left(\textrm{Cov}_\ell^{-1}\right)_{PP',QQ'}$ is the inverse of the covariance matrix between $C_\ell^{PP'}$ and $C_\ell^{QQ'}$ given by \cite{kable}
\begin{equation}
    \left(\textrm{Cov}_\ell\right)_{PP',QQ'}=\dfrac{1}{(2\ell+1)f_{\textrm{sky}}}\left(C_\ell^{PQ}C_\ell^{P'Q'}+C_\ell^{PQ'}C_\ell^{P'Q}\right)\:,
\end{equation}
with $f_{\textrm{sky}}$  being the sky fraction observed by the concerned instrument ($f_{\textrm{sky}}\to1$ for orbital satellite experiments). In our case, $PP'=QQ'=\mu T$ reduces the covariance matrix to a $1\times1$ form, yielding the Fisher matrix
\begin{equation} \label{fisher}
    F_{ij}\approx\sum_\ell(2\ell+1)f_{\textrm{sky}}\left(C_\ell^{\mu\mu,\textrm{N}}C_\ell^{TT,\textrm{fid}}\right)^{-1}\dfrac{\partial C_\ell^{\mu T}}{\partial\theta_i}\dfrac{\partial C_\ell^{\mu T}}{\partial\theta_j}\:.
\end{equation}
This approximation is justified as $C_\ell^{TT}$ is already measured to adequately high precision (i.e. reduces to the fiducial TT power spectrum $C_\ell^{TT,\textrm{fid}})$, and $C_\ell^{\mu\mu}$ is expected to be noise-dominated in our chosen framework (i.e. approximated by the experimental noise power spectrum $C_\ell^{\mu\mu,\textrm{N}})$ \cite{newwind}. From \eqref{clmutfin}, we have a one-dimensional parameter vector $\vec{\theta}=\{\theta_{\scriptscriptstyle{B}}\}$. Based on the Cram\'{e}r-Rao bound for a parametric model, $Q_\theta=\theta^2I_\theta$ represents the highest achievable SNR, where $I_\theta$ is the Fisher information \cite{quantum_info}. In the case of \eqref{clmutfin}, the SNR with respect to $\theta_{\scriptscriptstyle{B}}$ then boils down to
\begin{equation} \label{sn_interm}
    \left(\dfrac{S}{N}\right)\approx\sqrt{\sum_\ell(2\ell+1)f_{\textrm{sky}}\left(C_\ell^{\mu\mu,\textrm{N}}C_\ell^{TT,\textrm{fid}}\right)^{-1}\left(C_\ell^{\mu T}\right)^2}\:.
\end{equation}

For a given instrument, the SNR of the signal is then expected to be either higher or lower than for a purely BD scenario, corresponding respectively to either an enhancement or a suppression of the $C_\ell^{\mu T}$ amplitude (see figure \ref{fig:clmut}). In order to explicitly verify that and also to perform a quantitative estimation of the SNR, we choose a few proposed next-generation space-based CMB missions with the $C_\ell^{\mu\mu,\textrm{N}}$ noise templates given in table \ref{tab:noise_params}. Using \eqref{sn_interm}, the corresponding SNR values which we obtain for the combinations of the Bogolyubov parameters chosen for figure \ref{fig:clmut} are quoted in table \ref{tab:snr}.

\begin{table*}
\caption{\label{tab:noise_params}The experimental noise power spectrum $C_\ell^{\mu\mu,\textrm{N}}$ for a few next-generation satellite CMB missions aiming to probe spectral distortions. For a given frequency, the maximum observable multipole ($\ell_{\scriptscriptstyle{\rm max}}$) can be estimated from the angular Gaussian beam size at full-width-half-maximum. In case of relatively calibrated instruments, contributions from different frequency channels with different sensitivities need to be summed over \cite{Miyamoto:2013oua}.}
\begin{ruledtabular}
\begin{tabular}{ccccccc}
    {} & Experimental noise power spectrum of the $\mu\mu$-signal, i.e. $C_\ell^{\mu\mu,\textrm{N}}$ \cite{Miyamoto:2013oua,Chluba:2019nxa,Jeong:2019zaz} \\
    \hline \hline
    PIXIE \cite{pixie} & $1.3\times10^{-15}\times e^{(\ell/84)^2}$ \\ \hline
    Super-PIXIE \cite{super_pixie} & $4\pi\times10^{-18}\times e^{(\ell/84)^2}$ \\ \hline
    PICO \cite{NASAPICO:2019thw,pico} & $4.0\times10^{-19}\times e^{(\ell/2700)^2}$ \\ \hline
    LiteBIRD \cite{litebird} & $2.2\times10^{-18}\times e^{(\ell/135)^2}+1.8\times10^{-18}\times e^{(\ell/226)^2}$ \\
\end{tabular}
\end{ruledtabular}
\end{table*}

\begin{table*}
\caption{\label{tab:snr}Estimated signal-to-noise ratio of the $C_\ell^{\mu T}$ signal at the upcoming CMB missions of interest mentioned in table \ref{tab:noise_params}, corresponding to the representative combinations of Bogolyubov parameters chosen for figure \ref{fig:clmut}. The descending order of SNR corresponds to the display of figure \ref{fig:clmut}. The first six rows correspond to a BD scalar and an NBD gauge field sector (which enhances the signal) and the last four correspond to the opposite (which causes suppression), with the pure BD limit for both sectors sandwiched in between. Further enhancement of the SNR is restricted by the constraint $\widetilde{B}_\mu\lesssim27$ nG derived from the COBE/FIRAS+TRIS measurements of the $\mu$-monopole.}
\begin{ruledtabular}
\begin{tabular}{c c c c c c c c c}
        $\alpha_1$ & $\left(\dfrac{H}{M_{\rm Pl}}\right)\times10^{5}$ & $\gamma_2$ & $\widetilde{B}_\mu$ (in nG) & (S/N)\textsubscript{PIXIE} & (S/N)\textsubscript{Super-PIXIE} & (S/N)\textsubscript{PICO} & (S/N)\textsubscript{LiteBIRD} \\ 
        \hline \hline
        $1.0$ & $5.37$ & $60.0$ & $26.262$ & $4.591$ & $46.693$ & $262.900$ & $83.021$ \\ 
        $1.0$ & $5.37$ & $40.0$ & $17.506$ & $2.040$ & $20.749$ & $116.824$ & $36.892$ \\ 
        $1.0$ & $5.37$ & $20.0$ & $8.749$ & $0.510$ & $5.182$ & $29.179$ & $9.214$ \\ 
        $1.0$ & $5.37$ & $10.0$ & $4.366$ & $0.127$ & $1.291$ & $7.267$ & $2.295$ \\ 
        $1.0$ & $5.37$ & $5.0$ & $2.167$ & $0.031$ & $0.318$ & $1.789$ & $0.565$ \\ 
        $1.0$ & $5.37$ & $2.0$ & $0.817$ & $0.004$ & $0.045$ & $0.254$ & $0.080$ \\ \hline 
        \hline
        $1.0$ & $5.37$ & $1.0$ & $0.219$ & $3.188\times10^{-4}$ & $0.003$ & $0.018$ & $0.006$ \\ \hline 
        \hline
        $1.5$ & $2.07$ & $1.0$ & $0.084$ & $4.738\times10^{-5}$ & $4.819\times10^{-4}$ & $2.713\times10^{-3}$ & $8.568\times10^{-4}$ \\ 
        $2.0$ & $1.44$ & $1.0$ & $0.059$ & $2.293\times10^{-5}$ & $2.332\times10^{-4}$ & $1.313\times10^{-3}$ & $4.146\times10^{-4}$ \\ 
        $2.5$ & $1.12$ & $1.0$ & $0.046$ & $1.387\times10^{-5}$ & $1.411\times10^{-4}$ & $7.943\times10^{-4}$ & $2.508\times10^{-4}$ \\ 
        $3.0$ & $0.92$ & $1.0$ & $0.037$ & $9.359\times10^{-6}$ & $9.519\times10^{-5}$ & $5.359\times10^{-4}$ & $1.692\times10^{-4}$ \\ 
\end{tabular}
\end{ruledtabular}
\end{table*}

Considered together, figure \ref{fig:clmut} and table \ref{tab:snr} constitute the most important results of the present work on the observational front, exemplifying how detectable CMB signatures of primordial non-Gaussianity generated by magnetic fields may be modulated by the presence of non-standard initial vacua. Compared to the standard BD scenario, our results reflect significant differences in the amplitude of $C_\ell^{\mu T}$ sourced by the three-point $\langle\zeta BB\rangle$ correlation in presence of NBD initial conditions, which leads to important implications as far as its detectability is concerned. In case of an NBD scalar sector and a BD gauge field sector, the former entails an associated lowering of the Hubble parameter of inflation in order to keep the measured scalar power amplitude invariant, which, alongside the latter, results in an overall suppression of $C_\ell^{\mu T}$. Even for conservative values of the scalar NBD parameters, the resulting signal is found to quickly decline below the detectability threshold of next-generation CMB missions. A BD scalar sector, on the other hand, allows one to keep the Hubble parameter of inflation unchanged, while the accompanying NBD gauge field sector can significantly uplift the signal strength through enhancement of the magnetic field strength on the relevant $\mu$-distortion scales. 

While in our adopted framework it might still be difficult for PIXIE to observe a $C_\ell^{\mu T}$ signal enhanced to its maximum allowed limit via the presence of non-standard initial vacua, the prospects appear significantly better for the rest of the missions. As gleaned from table \ref{tab:snr}, the Super-PIXIE mission design could hit upon an SNR which might be, on an average, one order of magnitude higher than that achievable by PIXIE. For regions of the Bogolyubov parameter space which leads to magnetic field strengths close to the upper bound of $27$ nG, one may attain SNR $>10$ at Super-PIXIE which is usually required for clear detectability. As for the other two missions, i.e. PICO and LiteBIRD, the former shows markedly better prospect over the latter in terms of the SNR, which, in turn, is seen to be roughly twice of that realizable by Super-PIXIE in case of LiteBIRD. In particular, assuming such a specific scenario where a primordial scalar-magnetic bispectrum alone sources the detected signal, PICO and LiteBIRD should be able to constrain the effective nonlinearity parameter $\theta_B$ of the underlying model at around 10\% precision, with its associated one-sigma error being given by $\sigma(\theta_B)\equiv(\textrm{Det}( F_{ij}))^{-1/2}=\theta_B\times(S/N)^{-1}$. However, such possibilities are inherently model-dependent and we presently avoid any strong comments in this direction. Our analysis also explicitly confirms that the maximum observable multipole plays a sub-dominant role compared to the white noise level of $C_\ell^{\mu\mu,\textrm{N}}$ as far as the SNR is concerned, which renders the SNR at LiteBIRD roughly only three times lesser than that at PICO due to the instrumental noise specifications, in spite of the maximum observable multipole ($\ell_{\scriptscriptstyle{\rm max}}$) of LiteBIRD being a full order of magnitude smaller than that of PICO. Similarly, Super-PIXIE is expected to be a considerably superior probe compared to the baseline PIXIE design owing to its two orders of magnitude lower noise level, which lifts its SNR to the same order as that of LiteBIRD. 

Before winding up, let us briefly discuss a couple of other possible primordial origins of $C_\ell^{\mu T}$ that might pose as competitors to the scenario discussed above at future CMB missions.  The first notable one is the primordial scalar bispectrum $\langle \zeta\zeta\zeta\rangle$ parametrized by the scalar nonlinearity parameter $f_{\rm NL}$ \cite{newwind}, while the second one is due to a time-evolving scalar perturbation on superhorizon scales in the presence of PMFs \cite{Miyamoto:2013oua}. For canonical single-field inflationary models, one generically obtains $f_{\rm NL}\sim\mathcal{O}(\epsilon)\sim0.01$. Using the formalism of \cite{ganc_sloth} within our adopted framework of generic vacua, we then end up with $(S/N)\sim10^{-6}$ at PIXIE and $(S/N)\sim10^{-4}$ at the other three missions for the competing signal $C_{\ell,\:f_{\rm NL}}^{\mu T}$. For $C_{\ell,\:\zeta_{\rm ev}}^{\mu T}$ on the other hand, the SNR can reach only up to $\mathcal{O}(10^{-17})$ at PIXIE and $\mathcal{O}(10^{-13})$ at the other three missions for the case of the BD scalar and NBD magnetic sectors, whereas it is further suppressed by nearly four orders of magnitude for the reverse case. Thus, even in the presence of generic non-standard initial vacua, neither of these two signals is expected to possibly compete with the $C_\ell^{\mu T}$ sourced by the scalar-magnetic bispectrum with associated $(S/N)\gtrsim\mathcal{O}(1)$ at next-generation CMB missions, as demonstrated in the present article. 

A third possibility involves non-minimal scalar perturbations $\zeta_B$ sourced by the non-adiabatic magnetic pressure $\delta p_B=(4/3)\rho_B$ \cite{Caprini_2009,Suyama:2012wh,Nurmi:2013gpa}, that may have the form
\begin{equation}
    \zeta_B=-\int dt H(t)\delta p_B/(\rho_{\textrm{tot}}+p_{\textrm{tot}})\:.
\end{equation}
This may appear as a magnetic correction term on top of the vanilla curvature perturbation $\zeta$, and induce non-trivial scalar auto- and cross-bispectra amongst themselves. Out of such contributions, the $\langle\zeta\zeta_B\zeta_B\rangle$ correlator is expected to be a particularly strong source of non-Gaussianity \cite{Nurmi:2013gpa}, with an induced scalar nonlinearity parameter which is expected to be of order $f_{\rm NL}^{\rm (B)}\sim(\widetilde{B}_\mu/1\:\textrm{nG})^4$ in our currently adopted framework. For identical instrumental configurations, the SNR of a $C_\ell^{\mu T}$ signal potentially sourced by $\langle\zeta\zeta_B\zeta_B\rangle$, being proportional to $f_{\rm NL}^{\rm (B)}$, may overshoot the SNR of \eqref{clmutfin} by more than one order of magnitude for sufficiently large values of $\widetilde{B}_\mu$. While similar conclusions have been reached earlier by \cite{ganc_sloth}, we observe the same feature even for a model-dependent approach involving generic initial vacua. Moreover, owing to the separable form of \eqref{sqlim}, the $C_\ell^{\mu T}$ profiles sourced by $\langle\zeta BB\rangle$ and $\langle\zeta\zeta_B\zeta_B\rangle$ should have the same spectral shape even for arbitrary momentum profiles of the Bogolyubov coefficients. In other words, the extended parameter space opened up by generic initial conditions does not help us break the degeneracy between the leading order $C_\ell^{\mu T}$ signals corresponding to these two distinct primordial sources, which we highlight here as an important null result of our present work.

However, these conclusions remain quite strongly dependent on both the choice of the underlying inflationary model and post-inflationary physics of the magnetic field \cite{Kobayashi:2014sga}. In particular, there could exist different classes of models where NBD initial conditions may lead to a scale-dependent $f_{\rm NL}^{\rm (B)}$, which could then lead to $C_\ell^{\mu T}$ signals with distinct spectral shapes. Such possibilities need to be motivated by viable models of (pre)-inflationary physics, which require more rigorous investigations on a case-by-case basis. In view of our present scope, we defer these ideas to future studies.

\section{Discussions and conclusion} \label{sec:disc}

To summarize the key results, let us focus separately on the salient aspects of the two key parts of the present work. In the first half (sections \ref{sec:recap} \& \ref{sec:three_point_corr}), we have explicitly computed and studied the properties of the three-point correlation function between the curvature perturbation and primordial magnetic fields generated via direct gauge-inflaton coupling, under the assumption of generic initial vacua for both sectors. While the most general form of the correlator is rather unwieldy, the limiting triangular cases of the cross-bispectrum yield interesting insights. The squeezed limit, which is of immediate relevance to the second half of our work, allows a product form decomposition in terms of the power spectra of the long-wavelength curvature mode and the short-wavelength magnetic mode. While similar behavior has been observed  in the case of purely Bunch-Davies (BD) initial states, we reiterate that our results have widespread coverage of the parameter space for generic vacua with practically no additional assumption for both the scalar and the gauge field sectors, and thus are quite distinct from the BD case available in the literature. As such, the reduction of the squeezed cross-bispectrum to a two-point product form, even in arbitrary NBD regimes, constitutes a central and non-trivial result of the present work. Thus, even though the mathematical expression turns out to be a lookalike of the BD case, the non-trivial Bogolyubov coefficients for generic vacua allow additional degrees of freedom to play around with the numerical values, which may lead to quantitatively different and possibly interesting results as opposed to the BD scenario.  

On the other hand, when regulated by a finite past cut-off $\eta_0$ as the starting point of inflation, the flattened configuration (with a longer curvature mode) entails an $\mathcal{O}(\eta_0^2)$ NBD correction to the bispectrum in the limit of weak deviations from BD vacua. Such corrections are typical to flattened triangles for a variety of primordial three-point correlators in presence of excited initial states. While making contact with CMB observables, angular averaging should reduce the correction term by one order, which may still lead to potentially significant features that would be interesting to study in upcoming works. As for the two other triangular cases of interest, the equilateral and orthogonal limits are seen to contain additional logarithmic terms besides their usual BD limit counterparts, that might provide significant corrections at scales relevant to the CMB.

In the second half (sections \ref{sec:observable_sigs} \& \ref{sec:fisher}), we have proceeded to demonstrate how the presence of such generic vacua can in principle be of direct observational relevance in light of the CMB by exploiting the additional freedom as stated above. To that end, we have explored the possible effects of the Bogolyubov parameter space on a potential $C_\ell^{\mu T}$ correlation which may be sourced by the non-Gaussian scalar-magnetic cross-bispectrum in the squeezed limit, and have studied the observational prospects of the resulting signal at a few next-generation space-based CMB missions (i.e. PIXIE, Super-PIXIE, PICO, and LiteBIRD) via Fisher analysis. While the Bogolyubov coefficients can in general be complex-valued, we have focused in this work on their positive real-valued subset for the purpose of demonstrating its possible impact on the $C_\ell^{\mu T}$ amplitude relative to the BD case that stems from the presence of such NBD vacua.  For a BD scalar sector and an NBD gauge field sector, the strength of $C_\ell^{\mu T}$ can be significantly enhanced for similar values of the magnetic NBD parameters. The resulting SNR can be up to $\mathcal{O}(10)$ at Super-PIXIE and LiteBIRD, whose SNR values lie close to each other owing to similar instrumental noise levels in spite of considerable differences in the maximum observable multipole. More specifically, the SNR at LiteBIRD is expected to be roughly twice of that attainable by Super-PIXIE, which in turn should be able to hit upon an SNR one order of magnitude higher than the baseline PIXIE design. Finally, between PICO and LiteBIRD, the SNR at the former is found to be considerably higher compared to the latter for all the cases, which arguably renders PICO the most efficient mission among all four in terms of probing such a $C_\ell^{\mu T}$ signal. The overall trend observed among the obtained SNR values also confirms that the instrumental noise level is expected to dominate over the maximum observable multipole when it comes to estimating the SNR associated with $C_\ell^{\mu T}$. These findings are summarized in figure \ref{fig:clmut} and table \ref{tab:snr}, which constitute our key results which are of observational significance in the light of next-generation CMB missions. 

In section \ref{sec:fisher}, we have also considered a few other possible primordial sources of a $C_\ell^{\mu T}$ correlation, and briefly assessed their ability to source a competing signal in the regime of generic vacua. In particular, the magnetically induced scalar mode $\zeta_B$ sourced by non-adiabatic pressure of the PMFs may induce significant levels of non-Gaussianity via correlations of the form $\langle\zeta\zeta_B\zeta_B\rangle$, which might lead to a dominant $C_\ell^{\mu T}$ signal with identical spectral shape even under generic initial conditions. In alternative inflationary models where the presence of NBD vacua might produce a scale-dependent induced nonlinearity parameter $f_{\rm NL}^{\rm (B)}(k)$, this degeneracy in shape may possibly be broken. Such prospects, however, call for a comparative analysis across different fundamental models, which falls beyond the scope of the current study.

On the same note, the present work by no means offers an exhaustive exposition, as far as the nature of a three-point correlation between the primordial scalar mode and PMFs is concerned. For one thing, the results are heavily dependent on the choice of the underlying model, which determines the structure of the interaction Hamiltonian and therefore that of the correlator itself. Thus, from a theoretical point of view, it could be interesting to pursue similar works based on an EFT description of primordial magnetogenesis \cite{Kushwaha:2022bwy}, which might be able to provide insights into certain generic features of such correlators in a quasi model-independent manner (both in and without the presence of NBD vacua). At the same time, from an observational perspective, the model-dependent details can make both the $\langle \zeta BB\rangle$ and the $\langle \zeta \zeta_B\zeta_B\rangle$ correlators competent tools for discriminating among possible scenarios of PMF generation based on future data, which warrants further investigation for different models. To conclude, higher order correlations between primordial metric perturbations and cosmic magnetic fields are expected to source important cosmological observables across various epochs, which might play significant roles in refining our understanding of the very early history of our Universe in the era of precision cosmology. Beyond the present work, we intend to explore more of these interesting avenues in future studies.

\acknowledgments

The authors thank the anonymous referees for their constructive comments and scientific feedback, which helped in substantial improvement of the manuscript. We also thank Debarun Paul and Sourav Pal for fruitful discussions, and Rahul Shah and Debarun Paul for constructive feedback on the manuscript and for assistance with typesetting. The authors gratefully acknowledge the use of the publicly available code \textit{\href{https://github.com/cmbant/CAMB}{CAMB}}. AB thanks CSIR for financial support through Senior Research Fellowship (File no. 09/0093(13641)/2022-EMR-I). SP thanks the Department of Science and Technology, Govt. of India for partial support through Grant No. NMICPS/006/MD/2020-21. We acknowledge the computational facilities of the Indian Statistical Institute, Kolkata.
\\

\appendix

\section{Full expressions of \texorpdfstring{$\mathcal{J}_1$}{} and \texorpdfstring{$\mathcal{J}_2$}{} for generic initial vacua} \label{sec:appendix}

In this section, we provide the general expressions of the integrals $\mathcal{J}_1$ and $\mathcal{J}_2$, defined in \eqref{j1nbd} and \eqref{j2nbd} respectively, for arbitrary triangular configurations. 

\begin{widetext}
{
 \allowdisplaybreaks
  \begin{align}
        &\mathcal{J}_1=\:\frac{(\alpha_1+\beta_1) (\gamma_2+\delta_2) (\gamma_3+\delta_3)}{2(k_2 k_3)^{3/2}} \bigg[-\frac{\left(k_1^2+k_2^2\right) (\beta_1^* \gamma_3^* \delta_2^*-\alpha_1^* \gamma_2^* \delta_3^*)}{k_1+k_2-k_3}+\frac{\left(k_1^2+k_2^2\right) (\alpha_1^* \gamma_3^* \delta_2^*-\beta_1^* \gamma_2^* \delta_3^*)}{k_1-k_2+k_3} \nonumber \\
        &  +\frac{\left(k_1^2+k_2^2\right) (\beta_1^* \gamma_2^* \gamma_3^*-\alpha_1^* \delta_2^* \delta_3^*)}{-k_1+k_2+k_3}+\frac{\left(k_1^2+k_2^2\right) (\alpha_1^* \gamma_2^* \gamma_3^*-\beta_1^* \delta_2^* \delta_3^*)}{k_1+k_2+k_3}-\frac{k_1 k_2 (k_1+k_2) (\beta_1^* \gamma_3^* \delta_2^*-\alpha_1^* \gamma_2^* \delta_3^*)}{(k_1+k_2-k_3)^2} \nonumber \\
        & -\frac{k_1 k_2 (k_1-k_2) (\alpha_1^* \gamma_3^* \delta_2^*-\beta_1^* \gamma_2^* \delta_3^*)}{(k_1-k_2+k_3)^2}+\frac{k_1 k_2 (k_1-k_2) (\beta_1^* \gamma_2^* \gamma_3^*-\alpha_1^* \delta_2^* \delta_3^*)}{(-k_1+k_2+k_3)^2} \nonumber \\
        & +\frac{k_1 k_2 (k_1+k_2) (\alpha_1^* \gamma_2^* \gamma_3^*-\beta_1^* \delta_2^* \delta_3^*)}{(k_1+k_2+k_3)^2}+k_3 (\alpha_1^*+\beta_1^*) (\gamma_2^*+\delta_2^*) (\gamma_3^*-\delta_3^*)\bigg] + \textrm{h.c.}\:, \label{j1gen} \\
        & \nonumber \\
        &\mathcal{J}_2=\:\textrm{Im}\:\frac{(\gamma_2+\delta_2) (\gamma_3+\delta_3)}{2 (k_2 k_3)^{5/2}} \Bigg[-2 i (\alpha_1-\beta_1) (\alpha_1^*+\beta_1^*) (\gamma_2^*+\delta_2^*) (\gamma_3^*+\delta_3^*) k_1^3 \nonumber \\
        & -(\alpha_1+\beta_1) \pi  \bigg[12 \alpha_1^* \gamma_2^* \gamma_3^* k_1^3-12 \beta_1^* \gamma_2^* \gamma_3^* k_1^3+12 \alpha_1^* \gamma_3^* \delta_2^* k_1^3+12 \alpha_1^* \gamma_2^* \delta_3^* k_1^3 \nonumber \\
        & -\frac{i}{\pi } \Bigg\{\alpha_1^* \bigg\{\delta_3^* \bigg(\gamma_2^* \bigg(-3 (2 \gamma_E +3 i \pi +2 \ln (-\eta_I(k_1+k_2-k_3))) k_1^3+2 (k_1+k_2-k_3)^3 \nonumber \\
        & +6 (k_1+k_2-k_3) \left(k_1^2+(k_3-k_2) k_1+k_2 k_3\right) \nonumber \\
        & +\frac{2 k_2 k_3 \left(-3 (k_2-k_3) k_1^2+\left(-3 k_2^2+8 k_3 k_2-3 k_3^2\right) k_1+k_2 (k_2-k_3) k_3\right)}{(k_1+k_2-k_3)^2}\bigg) \nonumber \\
        & +\delta_2^* \bigg(-3 (2 \gamma_E +i \pi +2 \ln (-\eta_I(-k_1+k_2+k_3))) k_1^3+2 (k_1-k_2-k_3)^3 \nonumber \\
        & +6 (k_1-k_2-k_3) \left(k_1^2+(k_2+k_3) k_1-k_2 k_3\right) \nonumber \\
        & -\frac{2 k_2 k_3 \left(3 (k_2+k_3) k_1^2-\left(3 k_2^2+8 k_3 k_2+3 k_3^2\right) k_1+k_2 k_3 (k_2+k_3)\right)}{(-k_1+k_2+k_3)^2}\bigg)\bigg) \nonumber \\
        & +\gamma_3^* \bigg(\delta_2^* \bigg(-3 (2 \gamma_E +3 i \pi +2 \ln (-\eta_I(k_1-k_2+k_3))) k_1^3+2 (k_1-k_2+k_3)^3 \nonumber \\
        & +6 (k_1-k_2+k_3) \left(k_1^2+(k_2-k_3) k_1+k_2 k_3\right) \nonumber \\
        & -\frac{2 k_2 k_3 \left(-3 (k_2-k_3) k_1^2+\left(3 k_2^2-8 k_3 k_2+3 k_3^2\right) k_1+k_2 (k_2-k_3) k_3\right)}{(k_1-k_2+k_3)^2}\bigg) \nonumber \\
        & +2 \gamma_2^* \bigg(-\frac{3}{2} (2 \gamma_E +3 i \pi +2 \ln (-\eta_I(k_1+k_2+k_3))) k_1^3+(k_1+k_2+k_3)^3 \nonumber \\
        & +3 (k_1+k_2+k_3) \left(k_1^2-(k_2+k_3) k_1-k_2 k_3\right) \nonumber \\
        & +\frac{k_2 k_3 \left(3 (k_2+k_3) k_1^2+\left(3 k_2^2+8 k_3 k_2+3 k_3^2\right) k_1+k_2 k_3 (k_2+k_3)\right)}{(k_1+k_2+k_3)^2}\bigg)\bigg)\bigg\} \nonumber \\
        & -\beta_1^* \bigg\{8 \gamma_2^* \gamma_3^* k_1^3+8 \gamma_3^* \delta_2^* k_1^3+8 \gamma_2^* \delta_3^* k_1^3+8 \delta_2^* \delta_3^* k_1^3+3 \gamma_2^* \gamma_3^* \ln \left(\frac{i}{k_1-k_2-k_3}\right) k_1^3 \nonumber \\
        & +3 \gamma_3^* \delta_2^* \ln \left(\frac{i}{k_1+k_2-k_3}\right) k_1^3-3 \gamma_3^* \delta_2^* \ln (-i (k_1+k_2-k_3)) k_1^3 +3 \gamma_2^* \delta_3^* \ln \left(\frac{i}{k_1-k_2+k_3}\right) k_1^3 \nonumber \\
        & -3 \gamma_2^* \delta_3^* \ln (-i (k_1-k_2+k_3)) k_1^3 -3 \gamma_2^* \gamma_3^* \left(\frac{i \pi }{2}+\ln (-k_1+k_2+k_3)\right) k_1^3+3 \delta_2^* \delta_3^* \ln \left(\frac{i}{k_1+k_2+k_3}\right) k_1^3 \nonumber \\
        & -3 \delta_2^* \delta_3^* \ln (-i (k_1+k_2+k_3)) k_1^3-6 \gamma_2^* \gamma_3^* \ln (-\eta_I) k_1^3-6 \gamma_3^* \delta_2^* \ln (-\eta_I) k_1^3-6 \gamma_2^* \delta_3^* \ln (-\eta_I) k_1^3 \nonumber \\
        & -6 \delta_2^* \delta_3^* \ln (-\eta_I) k_1^3-6 \gamma_2^* \gamma_3^* \gamma  k_1^3-6 \gamma_3^* \delta_2^* \gamma  k_1^3-6 \gamma_2^* \delta_3^* \gamma  k_1^3-6 \delta_2^* \delta_3^* \gamma  k_1^3+6 \gamma_3^* \delta_2^* k_2 k_1^2-6 \gamma_2^* \delta_3^* k_2 k_1^2 \nonumber \\
        & +6 \delta_2^* \delta_3^* k_2 k_1^2-6 \gamma_2^* \gamma_3^* k_3 k_1^2-6 \gamma_3^* \delta_2^* k_3 k_1^2+6 \gamma_2^* \delta_3^* k_3 k_1^2+6 \delta_2^* \delta_3^* k_3 k_1^2-\frac{6 \gamma_3^* \delta_2^* k_2^2 k_3 k_1^2}{(k_1+k_2-k_3)^2} \nonumber \\
        & -\frac{6 \gamma_2^* \delta_3^* k_2 k_3^2 k_1^2}{(k_1-k_2+k_3)^2} -\frac{6 \gamma_2^* \gamma_3^* k_2 k_3^2 k_1^2}{(-k_1+k_2+k_3)^2}-\frac{6 \gamma_2^* \gamma_3^* k_2^2 k_3 k_1^2}{(-k_1+k_2+k_3)^2}+\frac{6 \gamma_3^* \delta_2^* k_2 k_3^2 k_1}{k_1+k_2-k_3}+\frac{8 \gamma_3^* \delta_2^* k_2^2 k_3^2 k_1}{(k_1+k_2-k_3)^2} \nonumber \\
        & -6 \gamma_2^* \gamma_3^* k_2 k_3 k_1  +6 \gamma_3^* \delta_2^* k_2 k_3 k_1+6 \gamma_2^* \delta_3^* k_2 k_3 k_1-6 \delta_2^* \delta_3^* k_2 k_3 k_1-\frac{6 \gamma_3^* \delta_2^* k_2^3 k_3 k_1}{(k_1+k_2-k_3)^2} \nonumber \\
        & -\frac{6 \gamma_2^* \delta_3^* k_2 k_3^3 k_1}{(k_1-k_2+k_3)^2} +\frac{8 \gamma_2^* \delta_3^* k_2^2 k_3^2 k_1}{(k_1-k_2+k_3)^2}+\frac{6 \gamma_2^* \gamma_3^* k_2 k_3^3 k_1}{(-k_1+k_2+k_3)^2}+\frac{14 \gamma_2^* \gamma_3^* k_2^2 k_3^2 k_1}{(-k_1+k_2+k_3)^2} \nonumber \\
        & +\frac{6 \delta_2^* \delta_3^* k_2 k_3^2 k_1}{k_1+k_2+k_3} +\frac{6 \delta_2^* \delta_3^* k_2^2 k_3 k_1}{k_1+k_2+k_3}+\frac{2 \delta_2^* \delta_3^* k_2^2 k_3^2 k_1}{(k_1+k_2+k_3)^2}-2 \gamma_2^* \gamma_3^* k_2^3+2 \gamma_3^* \delta_2^* k_2^3-2 \gamma_2^* \delta_3^* k_2^3 \nonumber \\
        & -2 \gamma_2^* \gamma_3^* k_3^3-2 \gamma_3^* \delta_2^* k_3^3 +2 \gamma_2^* \delta_3^* k_3^3+2 \delta_2^* \delta_3^* k_3^3+\frac{2 \gamma_3^* \delta_2^* k_2^2 k_3^2}{k_1+k_2-k_3}+\frac{2 \gamma_2^* \delta_3^* k_2^2 k_3^2}{k_1-k_2+k_3}-\frac{2 \gamma_2^* \gamma_3^* k_2^2 k_3^2}{-k_1+k_2+k_3} \nonumber \\
        & \left.\left.\left.\left. +\frac{2 \delta_2^* \delta_3^* k_2^2 k_3^2}{k_1+k_2+k_3}-6 \gamma_2^* \gamma_3^* k_2 k_1^2 +2 \delta_2^* \delta_3^* k_2^3 +\frac{6 \gamma_2^* \gamma_3^* k_2^3 k_3 k_1}{(-k_1+k_2+k_3)^2} +\frac{6 \gamma_2^* \delta_3^* k_2^2 k_3 k_1}{k_1-k_2+k_3} \right\}\right\}\right]\right] \label{j2gen}
\end{align}
}
\end{widetext}

\noindent The oscillatory parts of the integrals have been neglected while writing these expressions. As such, \eqref{j1gen} and \eqref{j2gen} are suitable for computing the squeezed, equilateral, and orthogonal limits of $\langle\zeta BB\rangle$ on the basis of \eqref{rbbbispec} and \eqref{I-J}. In the flattened configuration, the contribution of the oscillatory parts (given by complex exponential integrals) becomes non-negligible due to the assumption of a finite initial past, and regulates the apparent divergence of the bispectrum under NBD conditions. Due to computational complexity, we have not explicitly derived the most general forms of $\mathcal{J}_1$ and $\mathcal{J}_2$ by retaining the oscillatory integrals, having instead evaluated the flattened limit result separately in the weak NBD regime in section \ref{subsec:lim_triangles}. 

It can be checked that in the fully Bunch-Davies limit, both \eqref{j1gen} and \eqref{j2gen} reduce properly to the results obtained by \cite{Jain_Sloth} as expected.

\section{Results for equilateral and orthogonal shapes} \label{sec:appendix_2}

\subsubsection*{Equilateral limit}

In the \textit{equilateral} limit where $k_1=k_2=k_3=k$ and $\vec{k_i}.\vec{k_j}=k^2/2$, proceeding similarly as before, the cross-bispectrum reduces to
\begin{widetext}
{
 \allowdisplaybreaks
  \begin{align}
      &\mathcal{B}^{\rm (eq)}(k_1,k_2,k_3)=\frac{|\zeta_{k^*}|^2|A_{k^*}|^2}{9a(\eta_I)^4} k^2 \textrm{Re}\Bigg[(\gamma +\delta )^2 \Big[4 (\alpha +\beta ) \Big\{\alpha^* \left(17 \gamma^{*2}+54 \gamma^* \delta^*-27 \delta^{*2}\right) \nonumber \\
        & +\beta^* \left(27 \gamma^{*2}-54 \gamma^* \delta^*-17 \delta^{*2}\right)\Big\}+5 \Big\{27 (\alpha +\beta ) \Big(\ln (-3 k \eta_I) \left(\beta^* \delta^{*2}-\alpha^* \gamma^{*2}\right) \nonumber \\
        & +\ln (-k \eta_I) (\beta^* \gamma^* (\gamma^*+2 \delta^*)-\alpha^* \delta^* (2 \gamma^*+\delta^*))\Big)-27 \gamma _E (\alpha +\beta ) (\alpha^*-\beta^*) (\gamma^*+\delta^*)^2 \nonumber \\
        & +2 \gamma^* \Big(\gamma^* (47 \alpha  \alpha^*+56 \alpha^* \beta +9 \beta  \beta^*)+9 \delta^* (4 \alpha  \alpha^*+5 \alpha^* \beta -5 \alpha  \beta^*-4 \beta  \beta^*)\Big) \nonumber \\
        & +2 \delta^* \Big(9 \gamma^* (4 \alpha  \alpha^*+5 \alpha^* \beta -5 \alpha  \beta^*-4 \beta  \beta^*)-\delta^* (9 \alpha  \alpha^*+56 \alpha  \beta^*+47 \beta  \beta^*)\Big)\Big\}\Big]\Bigg] \nonumber \\
        & -\frac{15\pi}{2}\frac{|\zeta_{k^*}|^2|A_{k^*}|^2}{a(\eta_I)^4} k^2 \textrm{Im}\Big[(\alpha +\beta ) (\gamma +\delta )^2 \Big\{\alpha^* (\gamma^{*2}-\delta^{*2}+2\gamma^*\delta^*)+\beta^* \left(-\gamma^{*2}+2 \gamma^* \delta^*+\delta^{*2}\right)\Big\}\Big]\:,
  \end{align}
}
\end{widetext}

\noindent where $\gamma_E\:\approx0.5772$ is the Euler-Mascheroni constant, and the Bogolyubov coefficients are unlabeled as they all correspond to the same momentum $k$. While the $\ln(-3k\eta_I)$ term appears in the vanilla BD scenario, the presence of excited initial states in the scalar sector (non-zero $\beta$) and/or in the magnetic sector (non-zero $\delta$) gives rise to an additional $\ln(-k\eta_I)$ term, although our calculations are done entirely at the tree-level. This is an interesting feature, as under suitable numerical choices of the Bogolyubov parameters (including the total BD limit), terms of the form $\ln(-k_{o}\eta_I)\sim-60$ can potentially dominate the equilateral bispectrum, with $k_{o}$ corresponding to the observable CMB scale at present. Thus, it could be interesting to study the impact of such additional logarithmic contributions on cosmological observables.

\subsubsection*{Orthogonal limit}

The \textit{orthogonal} bispectrum $\mathcal{B}^{\rm (or)}(k_1,k_2,k_3)$, defined by $\vec{k_2}.\vec{k_3}=0$ and $k_1=\sqrt{2}k_2=\sqrt{2}k_3$, does not depend on the first integral $\mathcal{I}_1$ due to the form of \eqref{rbbbispec}, and has the explicit form
\begin{widetext}
{
 \allowdisplaybreaks
 \begin{align}
     &\mathcal{B}^{\rm (or)}(k_1,k_2,k_3)=-\frac{6 \sqrt{2}\pi}{a(\eta_I)^4} |\zeta_{k_1^*}|^2|A_{k_2^*}|^2 k_1^2 \textrm{Im}\Big[\left(\alpha _1+\beta _1\right) \left(\gamma _2+\delta _2\right)^2 \nonumber \\ 
        & \times \Big\{\alpha_1^* \Big(\gamma_2^* (\gamma_2^*+\delta_2^*)+\delta_2^* (\gamma_2^*-\delta_2^*)\Big)+\beta_1^* \Big(\gamma_2^* (\delta_2^*-\gamma_2^*)+\delta_2^* (\gamma_2^*+\delta_2^*)\Big)\Big\}\Big] \nonumber \\
        & -\frac{2}{a(\eta_I)^4} |\zeta_{k_1^*}|^2|A_{k_2^*}|^2 k_1^2 \textrm{Re}\Bigg[\left(\gamma _2+\delta _2\right) \left(\gamma _3+\delta _3\right) \Big[\alpha _1 \Big\{\alpha_1^* \Big(\gamma_2^* \big( (6 \sqrt{2} \gamma _E-10 \sqrt{2}-7 \nonumber \\
        & +6 \sqrt{2} \ln (\sqrt{2}+1))\gamma_2^*+2 \sqrt{2} \left(3 \gamma _E-5\right) \delta_2^*\big)+\delta_2^* \big(2 \sqrt{2} (3 \gamma _E-5) \gamma_2^*+\delta_2^* (6 \sqrt{2} \gamma _E-10 \sqrt{2} \nonumber \\
        & +7+6 \sqrt{2} \ln (\sqrt{2}-1))\big)\Big)+\beta_1^* \Big(\delta_2^* \big(2 \sqrt{2} (7-3 \gamma _E)\gamma_2^* +(-6 \sqrt{2} \gamma _E+14 \sqrt{2}+7 \nonumber \\
        & +3 \sqrt{2} \ln (\sqrt{2}-1)-3 \sqrt{2} \ln (\sqrt{2}+1))\delta_2^* \big)-\gamma_2^* \big( (6 \sqrt{2} \gamma _E-14 \sqrt{2}+7+6 \sqrt{2} \ln (\sqrt{2}-1))\gamma_2^* \nonumber \\
        & +2 \sqrt{2} (3 \gamma _E-7) \delta_2^*\big)\Big)\Big\}+\beta _1 \Big\{\alpha_1^* \Big(\gamma_2^* \big( (6 \sqrt{2} \gamma _E-14 \sqrt{2}-7+6 \sqrt{2} \ln (\sqrt{2}+1))\gamma_2^* \nonumber \\
        & +2 \sqrt{2} (3 \gamma _E-7) \delta_2^*\big)+\delta_2^* \big(2 \sqrt{2} (3 \gamma _E-7) \gamma_2^*+ (6 \sqrt{2} \gamma _E-14 \sqrt{2}+7+6 \sqrt{2} \ln (\sqrt{2}-1))\delta_2^*\big)\Big) \nonumber \\
        & +\beta_1^* \Big( \delta_2^*\big(2 \sqrt{2} (5-3 \gamma _E) \gamma_2^*+ (-6 \sqrt{2} \gamma _E+10 \sqrt{2}+7+3 \sqrt{2} \ln (\sqrt{2}-1)-3 \sqrt{2} \ln (\sqrt{2}+1))\delta_2^*\big) \nonumber \\
        & -\gamma_2^* \big( (6 \sqrt{2} \gamma _E-10 \sqrt{2}+7+6 \sqrt{2} \ln (\sqrt{2}-1))\gamma_2^*+2 \sqrt{2} (3 \gamma _E-5) \delta_2^*\big)\Big)\Big\} \nonumber \\
        & +6 \sqrt{2} (\alpha _1+\beta _1) (\alpha_1^*-\beta_1^*) (\gamma_2^*+\delta_2^*) (\gamma_2^*+\delta_2^*) \ln (-k_1\eta _I)\Big]\Bigg]\:.
 \end{align}
}
\end{widetext}

\noindent While rather cumbersome, this expression still allows a couple of immediate insights, both sharing qualitative similarities with the equilateral configuration. Firstly, the logarithmic term is likely to dominate for suitable combinations of the Bogolyubov coefficients, including in the total BD limit of both the scalar and gauge field sectors. Secondly, the imaginary part highlighted separately in the expression above is an artifact of complex-valued coefficients, as it vanishes both for the BD limit and for purely real coefficients.
\\

\section{Supplementary analysis in light of the tensor-to-scalar ratio} \label{sec:appendix_3}

Although the analysis done in the present article is sufficient to serve the major purpose, both for theoretical developments and for prospects analysis,  let us do a brief investigation on possible effects of the tensor-to-scalar ratio $r$ for the sake of completeness. The latest observations of Planck 2018 \cite{pl18_inf} and BICEP/Keck \cite{BICEP:2021xfz} jointly yield an upper limit of $r<0.036$ on the tensor-to-scalar ratio. Assuming a BD initial vacuum corresponding to a minimalistic primordial tensor sector for simplicity, one obtains the amplitude of the tensor power spectrum to be $A_t\approx(2/\pi^2)\left(H/M_{\rm Pl}\right)^2$. Thus, $r$ remains related to the first slow-roll parameter via the consistency relation $r\approx16\epsilon/|\alpha_1+\beta_1|^2$, which translates to the upper bound $\epsilon/|\alpha_1+\beta_1|^2<0.00225$. Since this is also precisely the combination which appears in the expression of $A_s$ in \eqref{A_s}, whose observed upper bound must remain fixed at $A_s=2.130\times10^{-9}$, one consequently obtains an upper bound on the Hubble parameter during inflation given by approximately $(H/M_{\rm Pl})<1.93\times10^{-5}$. 

As the slow-roll parameters are functions of the background potential, one may expect their values to be independent of the initial vacua of perturbations at leading order. Hence, we may choose to fix $\epsilon=0.00225$ which is the maximum allowed value for $\alpha_1=1$ and $\beta_1=0$. This choice is automatically consistent with the first inequality, as any $\alpha_1>1$ and $\beta_1=\sqrt{\alpha_1^2-1}$ subsequently yields a value smaller than the upper threshold.  Taking the second Hubble slow-roll parameter $\eta_{\scriptscriptstyle H}=\dot{\epsilon}/H\epsilon$  into account in the order of the Hankel function solution as $\nu\approx3/2+\epsilon+\eta_{\scriptscriptstyle H}/2$, the observed bounds on the scalar spectral index $n_s=0.9649\pm0.0042$ translate to the upper and lower bounds $|\eta_{\scriptscriptstyle H}|\in(0.0264,0.0393)$ via the relation $n_s=4-2\nu$. The constancy of $A_s$ then amounts to the constraint relation 
\begin{equation} \label{pl18bound_H_suppl}
    |\alpha_1+\beta_1|\left(\dfrac{H}{M_{\rm Pl}}\right)=\sqrt{8\pi^2A_s\:\epsilon}=1.93\times10^{-5}\:.
\end{equation} 
In other words, with the background slow-roll parameters kept fixed, an NBD scalar sector is associated with a correspondingly lowered value of $(H/M_{\rm Pl})$. Any value of $(H/M_{\rm Pl})$ so obtained is, once again, consistent with its previously deduced upper bound, which is saturated for $\alpha_1=1$ and $\beta_1=0$. At this point, the situation is qualitatively identical to \eqref{pl18bound_H}, albeit resulting in a slightly lower value of the Hubble parameter during inflation which is nevertheless of the same order as that obtained earlier. Since our main focus has been on forecasting, the precise numerical difference between \eqref{pl18bound_H} and \eqref{pl18bound_H_suppl} is largely inconsequential for our purpose. If one chooses to work with \eqref{pl18bound_H_suppl}, $C_\ell^{\mu T}$ may still be enhanced to its maximum allowed limit (corresponding to $\widetilde{B}_\mu\sim27$ nG) for $\gamma_2\sim400$ instead (vide figure \ref{fig:clmut} and table \ref{tab:snr}). Thus, we include this analysis for the sake of completeness, and as a proof of principle for the validity of the analysis in section \ref{sec:observable_sigs}.

\bibliography{biblio} 

\end{document}